\documentstyle[12pt,epsfig]{article} 
\def\bb{b \bar{b}} 
\def\cc{c \bar{c}} 
\def\qq{q \bar{q}} 
 
\def\cO{{\cal O}}

\begin{document}  
\vspace*{-2cm}  
\renewcommand{\thefootnote}{\fnsymbol{footnote}}  
\begin{flushright}  
hep-ph/9807332\\  
DTP/98/42\\  
July 1998\\  
\end{flushright}  
\vskip 65pt  
\begin{center}  
{\Large \bf All-Orders Resummation of Leading Logarithmic  \\[3truemm] 
 Contributions to Heavy Quark Production \\[5truemm] 
in  Polarized $\gamma\gamma$ Collisions} \\ 
\vspace{1.2cm} 
{\bf  
Michael Melles${}^1$\footnote{Michael.Melles@durham.ac.uk} and  
W.~James~Stirling${}^{1,2}$\footnote{W.J.Stirling@durham.ac.uk}  
}\\  
\vspace{10pt}  
{\sf 1) Department of Physics, University of Durham,  
Durham DH1 3LE, U.K.\\  
  
2) Department of Mathematical Sciences, University of Durham,  
Durham DH1 3LE, U.K.}  
  
\vspace{80pt}  
\begin{abstract}
The measurements of the $\gamma\gamma$ and $b \bar b$ partial decay widths
of an intermediate-mass Standard Model Higgs boson are among the most important
goals of a future photon linear collider. While in an initially polarized $J_z=0$
state the background process $\gamma\gamma \to b \bar b$ is suppressed by
$\frac{m^2_b}{s}$, radiative corrections remove this suppression and
are known to be very sizable at the one-loop level. In particular a new type
of purely hard double logarithmic (DL) correction can even make the cross section negative
at this order in perturbation theory. 
The second order term of the series is also known and
enters with a positive sign at the two loop level. From a theoretical as well
as a practical point of view it is clearly desirable to resum this series and
to know its high energy behavior. 
In this paper, we derive the series of these novel ``non-Sudakov'' logarithms
to all orders and calculate the high energy limit analytically. 
We also give explicit three loop corrections in the DL-approximation
as a check of our results and show that the three loop structure also reveals
the higher order behavior of mixed hard and soft double logarithms. 
We are thus able to resum the virtual DL-corrections to $\gamma\gamma \left( J_z=0 
\right) \to b \bar b$ to all orders. 
\end{abstract}
\end{center}  
\vskip12pt

\setcounter{footnote}{0}  
\renewcommand{\thefootnote}{\arabic{footnote}}  
  
\vfill  
\clearpage  
\setcounter{page}{1}  
\pagestyle{plain} 
 
\section{Introduction} 
 
Among the  many important physics processes which can be studied 
at a future high-energy photon linear collider (PLC) \cite{gin1,gin2,tel}, 
the production 
of Higgs bosons, $\gamma\gamma \to H$, stands out. 
In particular, identification of an intermediate-mass Standard Model 
 Higgs boson 
via its (dominant) $H \to \bb$ decay channels seems feasible 
and would provide an important measurement of  
the $\gamma\gamma$ and $b \bar b$ partial decay widths. 
 
The dominant (continuum) background to $\gamma\gamma\to H \to \bb$ 
comes from the Standard Model processes $\gamma\gamma \to \cc, \bb$. 
Various techniques can be used to suppress these. The most effective 
appears to be to {\it polarize} the initial--state photons: Higgs bosons 
are only produced from an initial state with $J_z = 0$, whereas 
the leading-order $\qq$ ($q=c,b$) backgrounds are dominantly produced 
from the $J_z = \pm 2$ initial state. More precisely, production 
from the $J_z = 0$ state is suppressed by $m_q^2/s$ \cite{bkso,jt1,jt2}.
Hence in the region of the Higgs resonance $\sigma_{\qq}(J_z=0)/ 
\sigma_{\qq} \sim m_q^2/M_H^2$ and the $\cc$ and $\bb$ backgrounds 
are heavily suppressed. 
 
Unfortunately, the $\cO(m_q^2)$ suppression of the $J_z=0$ background 
does not survive beyond leading order in QCD perturbation theory. 
In particular, there is no corresponding mass suppression of the 
$\cO(\alpha_s)$ $\gamma\gamma\to \qq g$  $J_z = 0$ amplitude squared. 
Although this gives rise notionally to three-jet configurations, in 
certain circumstances (for example, 
collinear $\gamma \to \qq$ splitting followed 
by $\gamma q \to q g $ Compton scattering) these can mimic the two-jet 
Higgs $\to \bb$ signal. Of course a combination of $b$-tagging 
of both jets and three-jet veto can {\it in principle} eliminate the 
bulk of this background, and one is left considering the perturbative 
corrections from virtual and soft real gluon emission. These have the 
same $\cO(m_q^2/s)$ suppression as the leading-order 
contribution\footnote{Unless otherwise stated, from now on all 
statements about amplitudes and cross sections will refer to the 
$J_z=0$ helicity projection.}  However the perturbation series 
now contains large double logarithmic corrections $\sim \alpha_s^n 
\log^{2n}(s/m_q^2)$ at each order $n$, which potentially can give 
sizable corrections. Apart from their phenomenological importance, 
these logarithms are of theoretical interest. They do not have 
an obvious `Sudakov' exponentiating behavior, arising from the fact 
that they originate in non-trivial regions of soft gluon phase space. 
Furthermore, for any particular virtual gluon diagram they are mixed 
up with the genuine infra-red logarithms $\sim \alpha_s^n 
\log^{n}(m_q^2/\lambda^2) \log^{n}(m_q^2/s)$, 
where $\lambda$ is a fictitious gluon mass 
introduced to control the soft divergences. 
 
In Ref.~\cite{fkm} the double logarithmic contributions were studied 
in considerable detail through two loops, i.e. up to and including 
$\cO(\alpha_s^2\log^4(s/m_q^2))$ contributions. Inspection of the one-and 
two-loop coefficients led to the claim that, for $m_q = m_b$ and 
$\sqrt{s} \simeq M_H$, the perturbation series was reasonably well 
approximated by the lowest two orders. However from both a 
theoretical and practical pont of view it is clearly desirable 
to have more information about the behavior of the double-logarithmic 
(DL) terms to all orders, in particular to see if resummation is possible. 
This is the goal of the present paper. As we shall see, a careful 
study of all the relevant two- and three-loop diagrams {\it does} 
reveal the pattern of the DL terms, and allows the all-orders result 
to be derived and resummed. 
 
In this paper we shall restrict ourselves to the theoretical study 
of the origin and calculation of the DL terms. In a future paper we will 
investigate their numerical effect on realistic PLC cross sections. 
We begin, in section~2, by deriving the leading order result for 
the $J_z = 0$ cross section, identifying the origin of the $\cO(m_q^2)$ 
suppression. In section~3 we discuss the structure of the one-loop 
corrections, show how the DL contributions arise, and compare these 
with the known one loop result \cite{jt1,jt2,fkm}.  
The main part of the paper is 
section~4, where we calculate all three-loop DL contributions 
and deduce the all-orders result. Section~5 contains a summary of 
our results.

\section{The Born Amplitude} 

\begin{center}
\begin{figure}
\centering
\epsfig{file=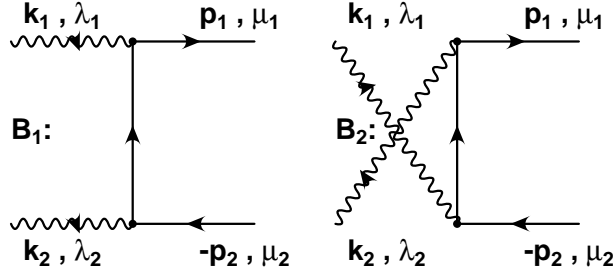,width=8cm}
\caption{The Born process in polarized $\gamma + \gamma \longrightarrow
q + \overline{q}$ with the notation of momenta used in the remainder of this
work. Keeping only the mass term in the numerator of the fermion 
propagator gives the correct
Born result only for $\theta \approx 0^o$. Details are given in the text.}
\label{fig:ppqqB}
\end{figure}
\end{center}

We begin with a rederivation of the underlying Born process. Recall that for
an initial state with $J_z=0$ the cross section vanishes in the massless limit.
In other words, the background to the standard Higgs production process via
a fermion loop and subsequent decay into a $b \overline{b}$ or 
$c \overline{c}$ final state is
suppressed by a factor of $\frac{m_q^2}{s}$. Higher order radiative 
(Bremsstrahlung) corrections remove this suppression in principle \cite{bkso}.
One therefore needs to keep at least terms of ${\cal O} \left( 
\frac{m_q}{\sqrt{s}} \right)$
in the calculation of the Born amplitudes shown in Fig. \ref{fig:ppqqB}.
We are also only interested in the region of large scattering angle $\theta$,
which is the dominant event topology of the Higgs signal \cite{jt1,fkm}.
  
Keeping only terms of ${\cal O} \left( \frac{m_q}{\sqrt{s}} \right)$, we find
it convenient to separate out the mass dependence of the spinors according to
\cite{ks}
\begin{eqnarray}
u^\mu (p) &=& \frac{ {\rlap/ p} + m}{ \sqrt{2p \cdot q}} | q, -\mu \rangle \label{eq:uspin} \\
v^\mu (p) &=& \langle q, -\mu | \frac{ - {\rlap/ p} + m}{ \sqrt{2p \cdot q}}  \label{eq:vspin}
\end{eqnarray}
where the spinors on the r.h.s. are massless and $q$ is an arbitrary reference
four-momentum with $q^2=0$. The separation of the mass terms then allows for the
convenient usage of the polarization vector \cite{ks}
\begin{eqnarray}
\varepsilon^\lambda_\nu (k,h) &\equiv& \frac{1}{\sqrt{4k \cdot h}} \langle k, \lambda | 
\gamma_\nu | h, \lambda \rangle \label{eq:epsmu} \\
\rlap/ \varepsilon^\lambda (k,h) &=& \frac{2}{\sqrt{4k \cdot h}} \left( | k, -\lambda \rangle
\langle h, -\lambda | + | h, \lambda \rangle \langle k, \lambda | \right) 
\label{eq:epsslash}
\end{eqnarray}
where here we denote the arbitrary massless reference momentum by $h$. In the
following we will also use simply $m$ for an arbitrary fermion mass, as the
results derived do not depend on any heavy quark approximation. It is understood
that for any application, the relevant fermion masses for the background to 
Higgs
production below the $W^+W^-$ threshold are the $b$- and $c$-quark masses.
The cross section of the latter is enhanced by a relative factor of $16$
from the different charges of the quarks.

Using the above notation we find for the Born amplitude ${\cal M}_{Born} \equiv B_1+B_2$
with an initial $J_z=0$ state:
\begin{equation}
{\cal M}_{Born} = \frac{8 m e^2 Q^2 \delta_{\lambda_1,\lambda_2} \delta_{\mu_1,\mu_2}
\delta_{\lambda_1,\mu_1}}{ \sqrt{s} (1-\beta^2 \cos^2 \theta) } \label{eq:ppqqB}
\end{equation}
with $\beta = \sqrt{1-\frac{4m^2}{s}}$, $e$ the electromagnetic coupling with
magnitude $Q$ and $\theta$ the scattering angle. The Kronecker-$\delta$'s
allow for a non-zero contribution only in case all four helicities are
the same. The result agrees with
Refs. \cite{fkm,jt1} up to ${\cal O} \left( \frac{m}{\sqrt{s}} \right)$ and displays the
somewhat surprising feature of a bosonic angular dependence
(as in Bhabha scattering for instance), with no $\theta$-terms in the numerator. 

This feature leads one to suspect that by keeping only the mass term of the
virtual fermion propagator, one should obtain Eq. \ref{eq:ppqqB} very easily.
A detailed analysis of this contribution, however, yields:
\begin{eqnarray}
{\cal M}^m_{Born} &=& \frac{4 m e^2 Q^2 \delta_{\lambda_1,\lambda_2} \delta_{\mu_1,\mu_2}
}{ s (1-\beta^2 \cos^2 \theta) } \left[
\langle k_1,-\lambda_1 | k_2, \lambda_1 \rangle 
(1+\beta^2 \cos^2 \theta) \delta_{\lambda_1,\mu_1} + \right. \nonumber \\
&& \left. \langle k_1,\lambda_1 | k_2, -\lambda_1 \rangle 
(1-\beta^2 \cos^2 \theta) \delta_{-\lambda_1,\mu_1} \right]
\label{eq:ppqqBm}
\end{eqnarray}

While it does not agree with the exact result in Eq. \ref{eq:ppqqB}, it is
also evident that for small angles and high energies both solutions do agree.
It was found in Ref. \cite{fkm} that this feature holds also through one and 
two loops if one is interested only in the leading logarithmic accuracy at
large angles. The reason for this seemingly contradictory behavior is that
while the angle between the outgoing fermion (antifermion) and the 
beamline is large, the angle between the incoming photon and the emitted
soft parton is small. Therefore, while it is wrong to keep only the $m$-term
in the propagator of the virtual fermion for the Born amplitude, it actually
{\it does} give the correct leading logarithmic structure of higher order
contributions with a soft fermion line.

For higher order corrections it is also convenient to use
the following expression for the $J_z=0$ ${\cal M}_{Born}$ amplitude:
\begin{equation}
{\cal M}_{Born} \approx - \frac{8 m e^2 Q^2}{s \; \sin^2 \theta}
{\varepsilon^{-\lambda}_1}_{\nu} (k_1,k_2) \; {\varepsilon^{-\lambda}_2}^{\nu}
(k_2,k_1) \; \langle p_1, \lambda | p_2, -\lambda \rangle \label{eq:ppqqBs}
\end{equation}

In the following we describe how to obtain the double logarithms
in the one loop approximation and introduce the techniques applied in the
later stages of the paper.

\section{One Loop Form Factors} \label{sec:1l}

At the one loop level, the only diagram that can contribute large double
logarithms is the box diagram with a virtual gluon connecting the outgoing
$q \overline{q}$ pair. While the general method of employing Sudakov 
parametrizations and integrating over the perpendicular components of 
soft loop
momenta is well known \cite{lan} and has been used extensively \cite{gglf,kl,ber,emr}, we still find
it useful to briefly review some basic properties. In section \ref{sec:res} 
we will also remark on some subtleties one encounters with higher order virtual
corrections.

Large logarithms occur only when there are small and large scales connected
in a Feynman diagram. While the particular Feynman diagram is gauge dependent,
the leading logarithmic contribution is not. In our case, the box diagram is
connected with vertex and self energy corrections to form a gauge
invariant cross section. However, in any gauge only the four-point diagram
can yield the desired pair of large logarithms. It is therefore independent
of the particular choice of gauge. A familiar and analogous example of the
same mechanism is the gauge invariance of the infra red $\frac{1}{\epsilon^2}$
term in the context of dimensional regularization for a massless box diagram.

In Fig. \ref{fig:1l} we list the four distinct topologies that allow a doubly
logarithmic contribution from the box diagram. The ``blob'' always denotes
a hard momentum flowing through the omitted propagator and in each case there
is a corresponding soft momentum flowing through the propagator facing the hard
blob. 

The feature which makes this process interesting and very different 
from the standard Sudakov
form factor is the fact that we have soft {\it fermion} contributions as well.
``Soft'' means in this context soft compared to the hard blob.

The Sudakov decomposition of loop momenta should be chosen such that 
the external four momenta joining the soft parton line are used as the base
vectors. 
\footnote{In higher orders this
is not always unambiguous, as we will discuss in section \ref{sec:res} and
can lead to technical difficulties.}
For the graphs of Fig. \ref{fig:1l}, 
however, we take:
\begin{center}
\begin{figure}
\centering
\epsfig{file=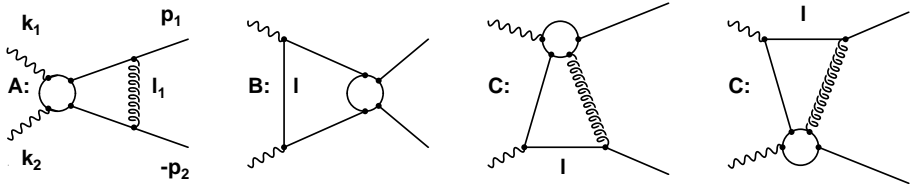,width=12cm}
\caption{The one loop soft and hard topologies 
 contributing to the 
 doubly logarithmic form factors. These graphs are
 obtained from the one loop box diagram. We denote the soft topology as 
 ${\cal A}$, the first hard topology as ${\cal B}$ and the second
 (third) as ${\cal C}$.} 
\label{fig:1l}
\end{figure}
\end{center}

\begin{eqnarray}
topology \;\; {\cal A} \;: l_1 &\equiv& \beta_1 \left( p_1 - \frac{m^2}{s} p_2 \right) + \alpha_1 \left( p_2 -
\frac{m^2}{s} p_1 \right) + {l_1}_\perp \label{eq:l1def} \\ 
topology \;\; {\cal B} \;: l \;\, &\equiv& \beta k_1 + \alpha k_2 
+ l_\perp \label{eq:Bldef} \\
topology \;\; {\cal C} \;: l \;\, &\equiv& \beta k_2 + \alpha \left( p_2 -
\frac{m^2}{s} k_2 \right) + l_\perp \label{eq:Cldef} 
\end{eqnarray}

At large angles and in the doubly logarithmic (DL) approximation, we have
\begin{equation}
2p_1 \cdot k_1 = 2p_2 \cdot k_2 \approx 2p_1 \cdot k_2=2p_2 \cdot k_1 \approx
2p_1 \cdot p_2 \approx 2k_1 \cdot k_2
=s \label{eq:DLapprox}
\end{equation}

The phase space for the large logarithms is given by
\begin{eqnarray}
1 &\gg& |\alpha|,|\beta| \gg |l^2_\perp/s|, 
\gg |m^2/s| \label{eq:DLappr} \\
1 &\gg& |\alpha_i|,|\beta_i| \gg
|{l_i}^2_\perp/s| \gg |\lambda^2/s| \label{eq:DL1appr}
\end{eqnarray}
where we already have multiple gluon insertions in mind. 
The four dimensional Minkowski measure is rewritten according to $d^4l=
d^2l_\perp d^2l_\parallel$, with
\begin{eqnarray}
d^2 l_\perp &=& |l_\perp| d |l_\perp| d \phi = \frac{1}{2} d l^2_\perp \; d \phi
= \pi d l^2_\perp \label{eq:perpm} \\
d^2 l_\parallel &=& | \partial (l_0,l_x) / \partial (\alpha,\beta)| \;d \alpha
\;d \beta \approx \frac{s}{2} \;d \alpha \; d \beta \label{eq:param}
\end{eqnarray}

The integrations
over the transverse momenta of the soft particles is performed by taking
half of the residues in the corresponding propagators:
\begin{eqnarray}
\frac{ d^4l}{l^2-m^2 +i \varepsilon} &=& \frac{s}{2} \frac{d \alpha \; d \beta
\; d^2 l_\perp}{s \alpha \beta + l^2_\perp -m^2 + i \varepsilon} 
\longrightarrow -i \pi^2 \frac{s}{2} d \alpha \; d \beta 
\; \Theta ( s \alpha \beta - m^2 ) \label{eq:softl} \\
\frac{ d^4l_i}{l_i^2-\lambda^2 +i \varepsilon} &=& \frac{s}{2} \frac{d \alpha_i \; d \beta_i
\; d^2 {l_i}_\perp}{s \alpha_i \beta_i + {l_i}^2_\perp -\lambda^2 + i \varepsilon}
\longrightarrow -i \pi^2 \frac{s}{2} d \alpha_i \; d \beta_i
\; \Theta ( s \alpha_i \beta_i - \lambda^2 ) \label{eq:softli} 
\end{eqnarray}

The DL-contribution of a particular Feynman diagram is thus given by
\begin{equation}
{\cal M}_k = {\cal M}_{Born} \; {\cal F}_k \label{eq:formfac}
\end{equation}
where the ${\cal F}_k$ are given by integrals over the remaining Sudakov parameters at the $n$-loop level:
\begin{eqnarray}
topology \;\; {\cal A} \;\;\;\;\;\;: \quad {\cal F}_k &=& \left( \frac{ \alpha_s}{2 \pi} 
\right)^n \prod^n_{i=1} \int^1_0 \int^1_0 \frac{d \alpha_i}{\alpha_i} \frac{d 
\beta_i}{\beta_i} \Theta_k \label{eq:SudA} \\
topology \;\; {\cal B} \;,\; {\cal C} \;: \quad {\cal F}_k &=& \left( \frac{ \alpha_s}{2 
\pi} \right)^n \int^1_0 \int^1_0 \frac{d \alpha}{\alpha}
\frac{d\beta}{\beta} \prod^{n-1}_{i=1} \int^1_0 \int^1_0 \frac{d \alpha_i}{\alpha_i}
\frac{d \beta_i}{\beta_i} \Theta_k \label{eq:NSudBC}
\end{eqnarray}

The factors $\Theta_k$ are specific to the given Feynman diagram and contain
its color factors and the restrictions on the occurring Sudakov variables
necessary to ensure that the matrix element has logarithmic behavior in
each of them. The range of integration is restricted by demanding that there
is at least one pole term on the opposite side of the contour of integration
from the other propagator terms,
since Cauchy's Theorem would otherwise give a vanishing contribution.
\begin{center}
\begin{figure}[t]
\centering
\epsfig{file=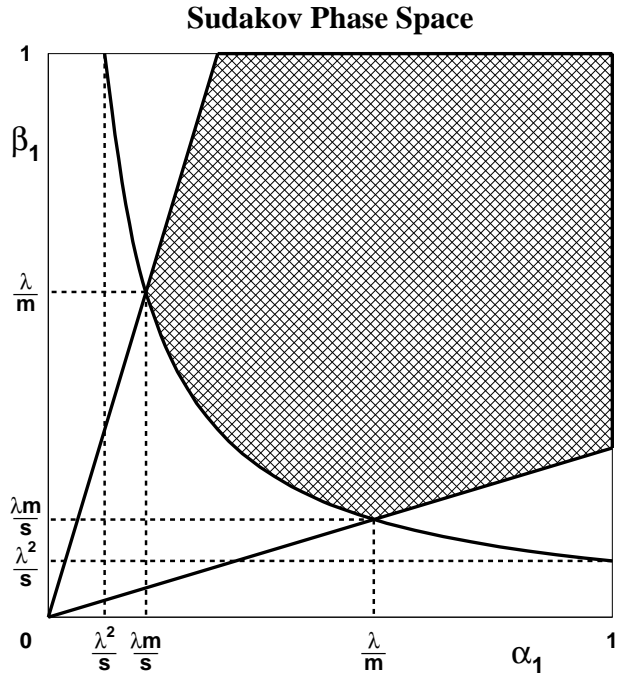,width=10cm}
\caption{The one loop soft Sudakov phase space of topology ${\cal A}$. 
The $\Theta$-functions given in Eq. \ref{eq:sff} are translated into integrals
covering the shaded area in the figure for the Sudakov parameters $\alpha_1$
and $\beta_1$ as defined in Eq. \ref{eq:l1def}. At higher orders, the
phase space becomes increasingly untrackable and even numerical integrations
need many iterations in order to converge.}
\label{fig:spsp}
\end{figure}
\end{center}

As a simple example, we now apply the above techniques to the one loop diagrams
of Fig. \ref{fig:1l}. For the standard infrared form factor of topology 
${\cal A}$ the Sudakov phase space is schematically given in Fig. \ref{fig:spsp}
and leads to the following result:

\begin{eqnarray}
{\cal F}_{\cal A} &=& - C_F \frac{\alpha_s}{2 \pi} \int^1_0 \int^1_0 
\frac{d \alpha_1}{\alpha_1} \frac{d \beta_1}
{\beta_1} \; \Theta \left( s \alpha_1 \beta_1 -\lambda^2 \right) \; \Theta \left(
\alpha_1 - \frac{m^2}{s} \beta_1 \right) \; \Theta \left( \beta_1 - \frac{m^2}{s}
\alpha_1 \right) \nonumber \\
&=& - C_F \frac{\alpha_s}{2 \pi} \left[ \int^\frac{\lambda}{m}_\frac{ 
\lambda^2}{s} \frac{d \beta_1}{\beta_1} \int^1_\frac{\lambda^2}{s \beta_1} 
\frac{d \alpha_1}{\alpha_1} + \int^1_\frac{\lambda}{m} \frac{d \beta_1}{\beta_1}
\int^1_{\frac{m^2}{s} \beta_1} \frac{d \alpha_1}{\alpha_1} 
-\int^\frac{\lambda m}{s}_\frac{\lambda^2}{s} \frac{d \beta_1}{\beta_1}
\int^1_\frac{\lambda^2}{s \beta_1} \frac{d \alpha_1}{\alpha_1} \right. \nonumber \\
&& \;\;\;\;\;\;\;\;\;\;\;\; \left. - \int^\frac{m^2}{s}_\frac{\lambda m}
{s} \frac{d \beta_1}{\beta_1} 
\int^1_{\frac{s}{m^2} \beta_1} \frac{d \alpha_1}{\alpha_1} \right] \nonumber \\
&=& - C_F \frac{\alpha_s}{2 \pi} \left( \frac{1}{2} \log^2 \frac{m^2}{s}
+ \log \frac{m^2}{s} \log \frac{\lambda^2}{m^2} \right) \label{eq:sff}
\end{eqnarray}

The result is identical to the usual DL-Sudakov form factor \cite{Sud}. For the
hard form factors of topologies ${\cal B}$ and ${\cal C}$ we have analogously:
\begin{eqnarray}
{\cal F} &\equiv& {\cal F}_{\cal B} = {\cal F}_{\cal C} = - C_F \frac{\alpha_s}{2 \pi}
\int^1_0 \int^1_0 \frac{d \alpha}{\alpha} \frac{d \beta}{\beta} \; \Theta \left(
s \alpha \beta -m^2 \right) \nonumber \\
&=& - C_F \frac{\alpha_s}{4 \pi} \log^2 \frac{m^2}{s} \label{eq:1lhff}
\end{eqnarray}

Taken together, and noting that topology ${\cal C}$ has a multiplicity factor $2$,
we thus have rederived the one loop structure of Refs. \cite{jt1,jt2} and 
\cite{fkm}
in the DL-approximation. The fact that the Born amplitude factors out is
trivial for topology ${\cal A}$. For topologies ${\cal B}$ and ${\cal C}$ the
exact numerator factorization was derived in Ref. \cite{fkm}. In the former, it is
connected
with the aforementioned purely bosonic angular dependence of ${\cal M}_{Born}$
of Eq. \ref{eq:ppqqB}, as it contains a hard subprocess $q + 
\overline{q} \longrightarrow q +\overline{q}$. In the latter, one needs to 
exploit the Dirac equation and gauge invariance in order to factor out
the hard subprocess $\gamma + q  \longrightarrow q + g$ and the Born amplitude.
Details are given in \cite{fkm}.

With these results, we can now proceed to investigate the higher order structure
of the leading logarithmic corrections from insertions of multiple gluons
in the basic contributions shown in Fig. \ref{fig:1l}. The next section gives
explicit results through three loops which enable us to arrive at an all orders
resummation of both hard as well as soft logarithmic corrections.

\section{All Orders Resummation} \label{sec:res}

In this section we derive the expressions for the leading logarithmic corrections
to all orders for the $J_z=0$ initial state $\gamma + \gamma \longrightarrow
q+ \overline{q}$ process. While it is possible to rigorously derive these
corrections to all orders for topologies ${\cal A}$ and ${\cal B}$, for 
topology ${\cal C}$ only the new hard ``non-Sudakov'' logarithms can be treated
analytically. For soft insertions we derive explicit three loop results and
show that essentially the same factorization structure emerges as in
${\cal B}$. This behavior is then extrapolated to all orders.

At this point we note that we have recalculated the virtual
two loop corrections presented in Ref. \cite{fkm} and agree with their results.
In the following, we therefore restrict ourselves to a discussion of the
three loop level. All resummed formulae were checked explicitly through one-,
two- and three-loops.

\subsection{The Sudakov Form Factor}

We begin with the Sudakov form factor of topology ${\cal A}$. At three loops,
the relevant Feynman diagrams are given in Fig. \ref{fig:A3ls}. The summation
is analogous to the soft vertex corrections where the exponentiation to
all orders for a color singlet final state was proved in 
Ref. \cite{ds} to leading order by employing an asymptotic Hamiltonian 
as well as the coherent
state formalism. Explicit three loop calculations
were performed in the leading logarithmic approximation in Ref. \cite{ct}
and related work can also be found in Refs. \cite{Sen,bu}.
For all orders corrections to topology ${\cal A}$ we thus find the 
familiar result:
\begin{equation}
{\cal F}^{soft} = \exp \left( {\cal F}_{\cal A} \right) \label{eq:softff}
\end{equation}

The non-Abelian color factors of diagrams with an ``Abelian'' topology are
canceled by the contributions of the purely non-Abelian topologies. In this
context it is worth noting that for the three gluon vertex coupling, there
is an effective factor of $-\frac{1}{2}$ stemming from two contributions
in numerator terms that are needed to obtain the required number of logarithms.
This is only the case if one has a denominator of the type $\frac{1}{(l-l_i)^2}$.
For one gluon very soft, there is a term $-2 l_i \cdot p$ and one $l_i
\cdot p$, $l_i$ being
the other (relatively) soft gluon momentum and $p$ an external hard fermion 
momentum. The Lorentz structure of the three gluon vertex is such that only
these two terms give DL-contributions (the third connects the same fermion
line). The effective soft contribution is therefore $-l \cdot p$, compared with
a factor $2p_\nu$ from the emission of a soft gluon off a fermion line.

\subsection{Resummation of Novel Hard Logarithms}

The interesting feature of the higher order corrections to the one-loop 
diagrams of Fig. \ref{fig:1l} is that there are novel purely hard 
``non-Sudakov''
logarithms due to the presence of the ``soft'' fermion line in topologies
${\cal B}$ and ${\cal C}$. In the case where a gluon insertion does not end on an external
(hard) fermion line, the soft fermion line serves essentially as a cutoff for
gluons that are soft compared to the hard blob. The diagrams are listed in
Figs. \ref{fig:B3l} and \ref{fig:C3l} and denoted by $h_i$, as they contain
no soft Sudakov contributions. Besides a different color factor, both
contributions are identical and lead to the same structure as in the soft case,
except for the last two integrations. They have the additional effect of
eliminating $\Theta \left( \alpha_i - \frac{m^2}{s} \beta_i \right)$ 
terms, which are very important in the soft case (see Fig. \ref{fig:spsp}).
The reason is simply that they are automatically fulfilled since
the soft fermion line induces a $\frac{m^2}{s}$ cutoff for the soft gluon
Sudakov variables. The integrations over the gluonic variables can thus
be performed easily and summing over $i$ gluon insertions
we have:
\begin{eqnarray}
{\cal F}^{hard} &\sim& \sum^\infty_{i=0} 
\left( \frac{\alpha_s}{2\pi} \right)^{i+1} 
\int^1_\frac{m^2}{s} \frac{d \beta}
{\beta} \int^1_\frac{m^2}{s \beta} \frac{ d \alpha}{\alpha} \frac{(-1)^{i+1}}
{i!} \log^i \beta \; \log^i \alpha \nonumber \\
&=& \sum^\infty_{i=0} \left( \frac{\alpha_s}{2\pi} \right)^{i+1} 
\frac{(-1)^i}{(i+1)!} \int^1_\frac{m^2}{s} \frac{d \beta}{\beta} \log^i \beta
\left( \log \frac{m^2}{s \beta} \right)^{i+1} \nonumber
\end{eqnarray}

\begin{center}
\begin{figure}[t]
\centering
\epsfig{file=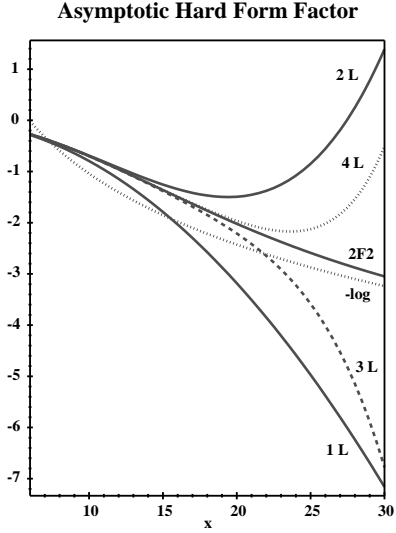,height=8cm}
\caption{A comparison of the exact summed result of Eq. \ref{eq:hff} with
the asymptotic form given in Eq. \ref{eq:ahff} and the first four terms in
the expansion. The value of the coupling was kept fixed at $\alpha_s=0.1$
and $x=\log \frac{s}{m^2}$. At a fixed order in perturbation theory the
high energy limit alternates. Only the exact and asymptotic form factors
can be seen to give the correct limit at high energies.}
\label{fig:hff}
\end{figure}
\end{center}

Using partial integration $i$-times (differentiating $\log^{i+1} \frac{m^2}{s
\beta}$, integrating $\frac{\log^i \beta}{\beta}$), we find:
\begin{center}
\begin{figure}[t]
\centering
\epsfig{file=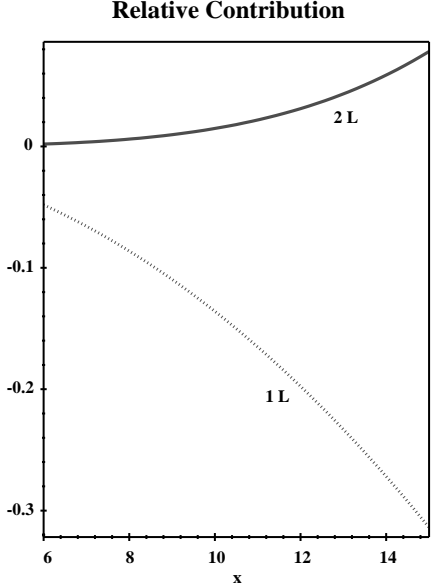,height=8cm}
\caption{The relative contribution of the one ($1 L$) and two ($2 L$) loop corrections
to the exact summed result of Eq. \ref{eq:hff}.
The value of the coupling was kept fixed at $\alpha_s=0.1$
and $x=\log \frac{s}{m^2}$. For the b-quark mass, $\sqrt{s}=100$ GeV corresponds
roughly to $x=6.2$, $\sqrt{s}=1$ TeV to $x=10.8$. In case of the c-quark,
$\sqrt{s}=100$ GeV corresponds roughly to $x=8.4$, $\sqrt{s}=1$ TeV to $x=13$.
The higher than second order corrections are thus important whenever the
experimental accuracy reaches the percentile level.}
\label{fig:rel}
\end{figure}
\end{center}

\begin{eqnarray}
{\cal F}^{hard} &\sim& 
\sum^\infty_{i=0} \left( \frac{\alpha_s}{2\pi} \right)^{i+1} 
\frac{(-1)^i \;(i+1)!}{(i+1)! \; (i+1)...2i} \int^1_\frac{m^2}{s} 
\frac{d \beta}{\beta} \log^{2i} \beta \;
\log \frac{m^2}{s \beta} \nonumber \\
&=& \sum^\infty_{i=0} \left( \frac{\alpha_s}{2\pi} \right)^{i+1}
\frac{(-1)^i \; \Gamma (i+1)}{\Gamma (2i+1)} \left[
\frac{1}{2i+1} \log^{2i+1} \beta \; \log \frac{m^2}{s} - \frac{1}{2i+2}
\log^{2i+2} \beta \right]^1_\frac{m^2}{s} \nonumber \\
&=& - \sum^\infty_{i=0} \left( \frac{\alpha_s}{2\pi} \right)^{i+1}
\frac{(-1)^i \; \Gamma (i+1)}{\Gamma (2i+3)} 
\log^{2i+2} \frac{m^2}{s} \nonumber \\
&=& - \frac{\alpha_s}{4\pi} \log^2 \frac{m^2}{s} \; _2F_2 (1,1;2,\frac{3}{2}; 
-\frac{\alpha_s}{8\pi} 
\log^2 \frac{m^2}{s})
\label{eq:hff}
\end{eqnarray}

In order to compare this result with the known one- and two-loop leading
logarithmic results we list them below explicitly and find agreement 
with Refs. \cite{jt1,jt2,fkm}:
\begin{eqnarray}
{\cal F}^{hard} &\sim& 
- \frac{1}{2} \left( \frac{\alpha_s}{2\pi} \right)
\log^{2} \frac{m^2}{s} 
+\frac{1}{24} \left( \frac{\alpha_s}{2\pi} \right)^2 \log^{4} \frac{m^2}{s}
-\frac{1}{360} \left( \frac{\alpha_s}{2\pi} \right)^3 \log^{6} \frac{m^2}{s}
\nonumber \\ &&
+ {\cal O} \left( \left( \frac{\alpha_s}{2\pi} \right)^4 \right)
\label{eq:exphff}
\end{eqnarray}

The asymptotic behavior of the hypergeometric function $_2F_2$ for $s 
\longrightarrow \infty$ can be obtained by using an integral representation
\cite{grad} and taking the leading pole term. In our case we find
\begin{eqnarray}
{\cal F}^{hard} &\sim& - \frac{\alpha_s}{4\pi} \log^2 \frac{m^2}{s} \; \Gamma \left( \frac{
3}{2} \right) \int^{i \infty}_{-i \infty} \frac{dt}{2 \pi i} \frac{ \Gamma^2
\left( 1+t \right) \Gamma \left( -t \right)}{\Gamma \left( 2 + t \right)
\Gamma \left( \frac{3}{2} + t \right)} \left( \frac{\alpha_s}{8\pi} \log^2 \frac{m^2}{s}
\right)^t \nonumber \\
&\stackrel{\textstyle \longrightarrow}{\scriptstyle s \to \infty}& - \log \left( \frac{\alpha_s}{\pi} \log^2 \frac{m^2}{s} \right)
\label{eq:ahff}
\end{eqnarray}

The last line follows from the residue of the integral representation at 
$t=-1$, closing the contour in the negative $t$-plane. All higher negative
pole contributions are suppressed by at least a factor of $\frac{1}{
\log^2 \frac{m^2}{s}}$. With the $\frac{m}{\sqrt{s}}$ suppression contained
in the Born amplitude (\ref{eq:ppqqB}), the high energy limit is well behaved.
\begin{center}
\begin{figure}[t]
\centering
\epsfig{file=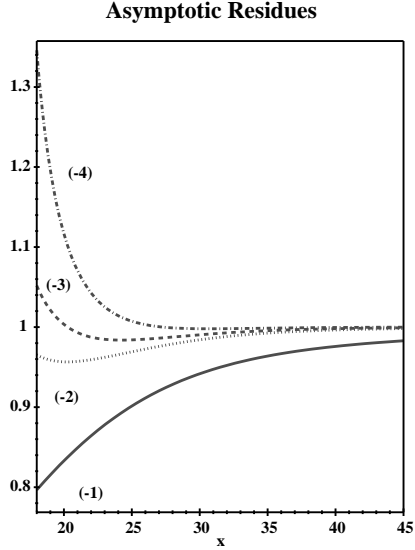,height=8cm}
\caption{A ratio of the first four residues of the integral
representation of Eq. \ref{eq:ihff} with the exact result of Eq. \ref{eq:hff}.
The value of the coupling was kept fixed at $\alpha_s=0.1$
and $x=\log \frac{s}{m^2}$. At lower energies the $\frac{1}{\log^2 
\frac{m^2}{s}}$ divergences of the higher order residues is clearly visible.
At high energies, however, they can be seen to give the correct limit.}
\label{fig:asres}
\end{figure}
\end{center}

It is also possible to reconstruct the asymptotic series based on the 
above integral representation. The higher order poles are simple ones
and can thus be evaluated easily. We find
\begin{eqnarray}
{\cal F}^{hard} &\sim& - \frac{\alpha_s}{4\pi} \log^2 \frac{m^2}{s} \; \Gamma \left( \frac{
3}{2} \right) \int^{i \infty}_{-i \infty} \frac{dt}{2 \pi i} \frac{ \Gamma^2
\left( 1+t \right) \Gamma \left( -t \right)}{\Gamma \left( 2 + t \right)
\Gamma \left( \frac{3}{2} + t \right)} \left( \frac{\alpha_s}{8\pi} \log^2 \frac{m^2}{s}
\right)^t \nonumber \\
&=& - \gamma - \log \left( \frac{\alpha_s}{2 \pi} \log^2 \frac{m^2}{s} \right)
+ \sum^\infty_{n=2} \frac{2^n}{(n-1)\pi} \frac{\Gamma \left( \frac{3}{2} \right)
\Gamma \left( n - \frac{1}{2} \right) }{ \left( \frac{\alpha_s}{4 \pi}
\log^2 \frac{m^2}{s} \right)^{n-1}}
\label{eq:ihff}
\end{eqnarray}

Although it possesses the ominous factorial growth of the coefficients in
the series for the higher order residue contributions, the equivalent
original expansion in terms of the confluent hypergeometric function in
Eq. \ref{eq:hff} means that it is Borel summable and that the high energy
limit in Eq. \ref{eq:ahff} does indeed give the correct behavior for $
s \longrightarrow \infty$. 
Fig. \ref{fig:hff} contains a comparison of the exact hard form factor
of Eq. \ref{eq:hff} and its asymptotic logarithmic divergence given by
the first term in Eq. \ref{eq:ihff}. They can be seen to agree in the high
energy limit. In addition the figure demonstrates that at a fixed order
in perturbation theory, the result alternates between $\pm \infty$ for
high energies.

The numerical effect of the higher order terms is shown in Fig. \ref{fig:rel}
where the all orders hard form factor is compared to the one- and two-loop
corrections at realistic future collider energies. As explained in the figure,
only if the experimental accuracy reaches the percentile level are
higher than second order contributions important. The relative size of the 
all orders correction
is larger for $m=m_c$ than for the b-quark mass.

A more detailed comparison of the behavior of the higher order residues
is given in Fig. \ref{fig:asres}. The first four pole contributions are
plotted against the exact hard form factor and agree with Eq. \ref{eq:hff}
in the asymptotic regime. At lower energies the $\frac{1}{ \left( \log^2 
\frac{m^2}{s} \right)^n}$ behavior of the first three simple residues
of Eq. \ref{eq:ihff} is clearly visible.

In order to check this derivation explicitly at the three loop level, we have
calculated the corresponding diagrams denoted by $h$ in Figs. \ref{fig:B3l}
and \ref{fig:C3l} and find the following results (modulo color factors and
chosing for the Sudakov decomposition $\left\{ k_1, k_2 \right\}$ for topology
${\cal B}$ and $\left\{ k_2, \left( p_2 - \frac{m^2}{s} k_2 \right) \right\}$ 
for topology ${\cal C}$):
\begin{eqnarray}
\Theta_{h_1} &\sim& \Theta \left( s \alpha \beta - m^2 \right) \Theta \left( \alpha_2
- \alpha_1 \right) \Theta \left( \alpha_1- \alpha \right) \Theta \left( 
\beta_2 - \beta_1 \right) \Theta \left( \beta_1 - \beta \right) \label{eq:3lh1}
\\ \Theta_{h_2} &\sim&
\Theta \left( s \alpha \beta - m^2 \right) \Theta \left( \alpha_1
- \alpha_2 \right) \Theta \left( \alpha_2- \alpha \right) \Theta \left( 
\beta_2 - \beta_1 \right) \Theta \left( \beta_1 - \beta \right) \label{eq:3lh2}
\end{eqnarray}

The remaining terms $h_3$ and $h_4$ in each figure cancel the non-Abelian
color factor stemming from the crossed box insertion $h_2$, i.e. 
$h_3=h_4 \sim \frac{h_2}{2}$. It can easily be seen
from the explicit form in Eqs. \ref{eq:3lh1} and \ref{eq:3lh2} that the
contributions of the gluon insertions are cut off effectively by the fermion
line Sudakov variables for which $\frac{m^2}{s}$ is the minimum value
attainable.

\begin{center}
\begin{figure}[t]
\centering
\epsfig{file=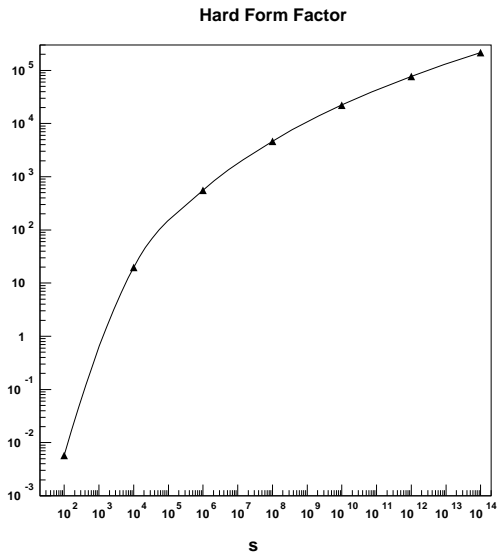,width=8cm}
\caption{A comparison of the third term of the series expansion in Eq.
\ref{eq:hff} (line) and the explicit three loop results given in
Eqs. \ref{eq:3lh1} and \ref{eq:3lh2} (triangles). For the three loop
data we used $10^6$ evaluations for each of the $100$ iterations of the
Monte Carlo integrator VEGAS \cite{veg}.
$m_b = 4.5$ GeV was used in both results. The absolute values are plotted
and the agreement is below the percent
level. Color factors and the coupling constant are omitted.}
\label{fig:3lh}
\end{figure}
\end{center}

Fig. \ref{fig:3lh} contains a comparison of the explicit three loop results given
by $h_1+h_2$ and the third term in the series \ref{eq:hff}.
The agreement is below the percentile level.

For completeness, we give the form factors explicitly for the two topologies
${\cal B}$ and ${\cal C}$ below, including the coupling and their differing 
color structure:

\begin{eqnarray}
{\cal F}^{hard}_{\cal B} 
&=& {\cal F} \;\; 
_2F_2 (1,1;2,\frac{3}{2}; \frac{1}{2} 
{\cal F} ) \label{eq:Bhff} \\
{\cal F}^{hard}_{\cal C} 
&=& {\cal F} \;\; 
_2F_2 (1,1;2,\frac{3}{2}; \frac{C_A}{4 C_F} 
{\cal F} ) \label{eq:Chff} 
\end{eqnarray}

where ${\cal F}$ is the one-loop hard form factor given in Eq. \ref{eq:1lhff}.

\subsection{Mixed Form Factors}

While the novel hard logarithms of the previous section are the most interesting
new feature of the $J_z=0$ $\gamma + \gamma \longrightarrow q + \overline{q}$
process, technically the mixed hard and soft contributions are a considerable
challenge. In fact for topology ${\cal C}$ we are only able to show numerically
through three loops that the same structure emerges as for topology ${\cal B}$.

\subsubsection{Topology ${\cal B}$}

We therefore start with the diagrams given in Fig. \ref{fig:B3l} 
for topology ${\cal B}$ which contain
a soft (s) contribution. The diagram $s_5$ is evidently given by the
product of the two-loop purely hard and the one-loop Sudakov form factors.
Also for the sum of diagrams $s_1+ ... + s_4$, it can easily be seen that we
again have a factorization, this time the purely hard one-loop times the
two-loop Sudakov form factor. 
\begin{center}
\begin{figure}
\centering
\epsfig{file=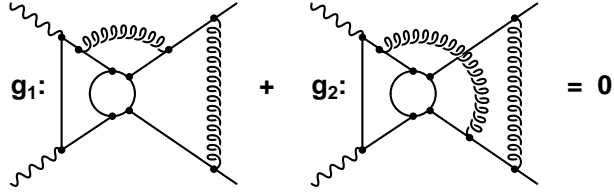,width=8cm}
\caption{A typical pair of three loop diagrams for topology ${\cal B}$
which vanishes in the DL-approximation when both
contributions of a gluon radiated off a loop left of the hard blob couples
to the outgoing fermion ( $g_1$ ) and the one coupling to the outgoing 
anti-fermion ( $g_2$ ) line are added.
The origin of this cancellation, which proceeds to all orders, is the spinor
line minus sign of the anti-fermion. Details are discussed in the text.}
\label{fig:B3lc}
\end{figure}
\end{center}

The nice feature of this topology is that
corrections connecting the original one-loop hard form factor with gluon
insertions between the outgoing fermions vanish in pairs. The reason is simply
a relative spinor line minus sign when a gluon starting from a fermion line
of the original one-loop topology is connected with one of the final state
fermions. One always has in this case an analogous correction which ends on the
anti-fermion line, thus yielding a relative minus sign. Fig. \ref{fig:B3lc}
indicates the cancellation in pairs. In this particular case, for instance,
we find (chosing the Sudakov parametrization $\left\{ k_1, \left( p_1- \frac{
m^2}{s} k_1 \right) \right\}$ for $g_1$ and $\left\{ k_1, \left( p_2- \frac{
m^2}{s} k_1 \right) \right\}$ for $g_2$ with respect to the gluon going
across the hard blob):

\begin{eqnarray}
\Theta_{g_1} &\sim& - \Theta \left( s \alpha \beta - m^2 \right) 
\Theta \left( s \alpha_1 \beta_1 - \lambda^2 \right)
\Theta \left( s \alpha_2 \beta_2 - \lambda^2 \right)
\Theta \left( \alpha_1 - \frac{m^2}{s} \beta_1 \right) \nonumber \\ &\times&
\Theta \left( \beta_1 - \frac{m^2}{s} \alpha_1 \right)
\Theta \left( \beta_2 - \frac{m^2}{s} \left(\alpha_2 - \beta_1 \right) \right)
\Theta \left( \alpha_2 - \alpha \right) \Theta \left( \beta_2 - \alpha_1 
\right) \\
\Theta_{g_2} &\sim& \Theta \left( s \alpha \beta - m^2 \right) 
\Theta \left( s \alpha_1 \beta_1 - \lambda^2 \right)
\Theta \left( s \alpha_2 \beta_2 - \lambda^2 \right)
\Theta \left( \alpha_1 - \frac{m^2}{s} \beta_1 \right) \nonumber \\ &\times&
\Theta \left( \beta_1 - \frac{m^2}{s} \alpha_1 \right)
\Theta \left( \beta_2 - \frac{m^2}{s} \left(\alpha_2 - \alpha_1 \right) \right)
\Theta \left( \alpha_2 - \alpha \right) \Theta \left( \beta_2 - \beta_1 \right)
\end{eqnarray}

The $g_2$ contribution is minus that of $g_1$ after the replacement
$\alpha_1 \longleftrightarrow \beta_1$ has been made. The two diagrams therefore
cancel each other.
This feature holds to
all orders and thus leads to a product of the hard form factor in Eq.
\ref{eq:Bhff} and the Sudakov form factor in Eq. \ref{eq:softff}.
In other words we have
\begin{equation}
{\cal F}^{hs}_{\cal B} =
{\cal F} \;\; 
_2F_2 (1,1;2,\frac{3}{2}; \frac{1}{2}  
{\cal F} ) \; \left[  \; \exp \left( {\cal F}_{\cal A} \right) -1
\right] \label{eq:Bhsff}
\end{equation}
where we have again used the notation introduced in section \ref{sec:1l} for
the one-loop form factors. $hs$ refers to the mixed hard (h) and soft (s)
nature of the form factor.

\subsubsection{Topology ${\cal C}$}

\begin{center}
\begin{figure}[t]
\centering
\epsfig{file=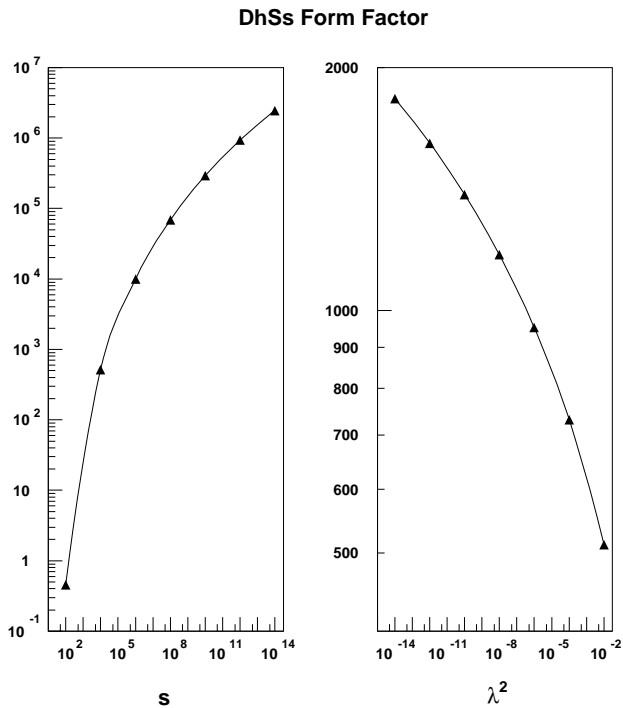,height=10cm}
\caption{A comparison of the third term of the series expansion in Eq.
\ref{eq:Chsff} (line) with a double hard and single soft form factor (DhSs)
and the explicit three loop results given in
appendix \ref{sec:hhs} (triangles).
The plot on the left keeps $\lambda^2=10^{-2}$ fixed, the one on the right
has $s=(100$ GeV$)^2$ for each data point. For the three loop
results we used $10^6$ evaluations for each of the $100$ iterations and
$m_b = 4.5$ GeV was used in both plots. The absolute values are plotted
and the agreement is below the percent
level. Color factors and the coupling constant are omitted.}
\label{fig:C3lhhs}
\end{figure}
\end{center}

For the diagrams given in Figs. \ref{fig:C3l} and \ref{fig:C3lsna} there is no
obvious way of generalizing a cancellation between diagrams to all orders.
We therefore chose to calculate these corrections explicitly through three
loops and compare with the ``expected'' analogous result found for topology
${\cal B}$ in Eq. \ref{eq:Bhsff}. The results are given in appendix 
\ref{sec:C3l}. Fig. \ref{fig:C3lhhs} shows a comparison of the contribution
from the second term of the hard form factor from Eq. \ref{eq:Chff} multiplied
by the one-loop soft form factor of Eq. \ref{eq:sff} with the three loop
corrections given by $s_{13}+...s_{20}$ in Fig. \ref{fig:C3l}. Only the diagrams
$s_{13}+s_{14}+s_{15}$ actually contribute, the others are either zero
($s_{19}$ and $s_{20}$) or cancel in the sum ($s_{16}+s_{17}+s_{18}=0$) in the DL-approximation.
The leading logarithmic
behavior is shown to agree for both $\lambda \to 0$ as well as for
$s \to \infty$, keeping $m=m_b$ fixed.
The agreement is better than $1\%$ and improves for large values of the
logarithms.
\begin{center}
\begin{figure}[t]
\centering
\epsfig{file=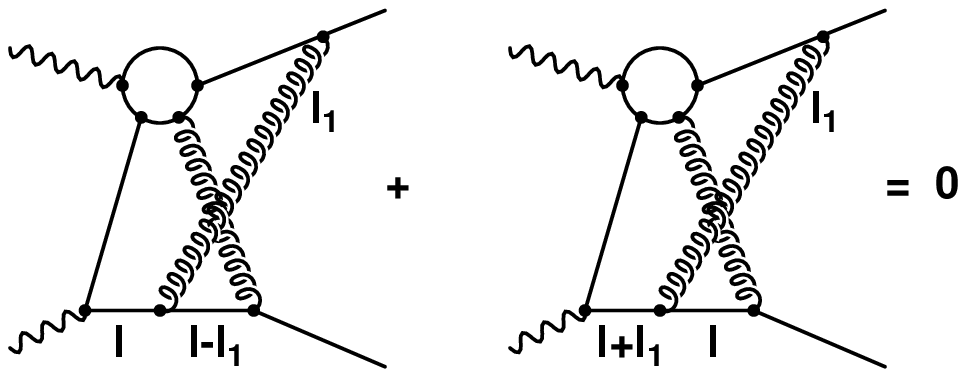,width=8cm}
\caption{A two loop diagram for topology ${\cal C}$ which vanishes 
in the DL-approximation when both
contributions of the soft fermion momentum $l$ to the left and right of the
gluon insertion are treated properly. As there is no ``obvious'' choice for
the Sudakov parametrization of the gluon momentum, technical difficulties
can arise. Details are discussed in the text.}
\label{fig:2l}
\end{figure}
\end{center}

For the other mixed hard and soft contribution at the three loop level, namely
the one-loop hard form factor in Eq. \ref{eq:1lhff} multiplied by the second term
in the Sudakov exponential of Eq. \ref{eq:softff}, many more diagrams contribute.
The remaining soft diagrams of Figs. \ref{fig:C3l} and \ref{fig:C3lsna} are not
only more numerous but also contain technical subtleties worth mentioning.

In particular, diagrams containing a soft gluon radiating off the ``soft''
fermion line facing the hard blob need careful consideration. In those cases
one has two (for diagrams $s_{11}$ and $s_{12}$ of Fig. \ref{fig:C3l} even
three) regions in the phase space contributing to the DL-corrections, 
corresponding to either side of the ``soft'' fermion line going on shell.

For instance the two loop diagram in Fig. \ref{fig:2l} possesses this ambiguity.
It actually does not give a contribution in the DL-approximation due to an
asymmetry of the diagram with respect to the external momenta $\left\{ k_2,p_2
\right\}$, while the large variables (see Eq. \ref{eq:DLapprox}) do not differ 
in the DL-approximation if we make the 
replacement $k_2 \longleftrightarrow p_2$. The three loop diagrams $s_{19}$ 
and $s_{20}$ in
Fig. \ref{fig:C3l} vanishes for the same reason. The easiest way to see this
on a technical level is therefore to parametrize $l_1$ choosing $\left\{ k_2,p_2
\right\}$ for the Sudakov decomposition. Expanding around the left side
of the ``soft'' fermion line in Fig. \ref{fig:2l} we find:

\begin{eqnarray}
\Theta_1 &\sim& \Theta \left( s \alpha \beta - m^2 \right) \; \left[ \Theta \left(
\beta_1-\alpha_1 \right) \Theta \left( \beta-\beta_1 \right) \Theta \left(
\alpha_1 \beta - \alpha \beta_1 -\alpha \beta - \alpha_1 \beta_1 \right) \right.
\nonumber \\
&& + \left. \Theta \left(
\alpha_1-\beta_1 \right) \Theta \left( \beta-\beta_1 \right) \Theta \left(
\alpha \beta_1- \alpha_1 \beta -\alpha \beta - \alpha_1 \beta_1 \right) \right]
\end{eqnarray}

Shifting the fermion loop momentum, effectively $l \longrightarrow l + l_1$, and
expanding around the right side of the gluon insertion gives analogously:
\begin{eqnarray}
\Theta_2 &\sim& -\Theta \left( s \alpha \beta - m^2 \right) \; \left[ \Theta \left(
\alpha_1 - \beta_1 \right) \Theta \left( \alpha-\alpha_1 \right) \Theta \left(
\alpha \beta_1 - \alpha_1 \beta -\alpha \beta - \alpha_1 \beta_1 \right) \right.
\nonumber \\
&& + \left. \Theta \left(
\beta_1-\alpha_1 \right) \Theta \left( \alpha-\alpha_1 \right) \Theta \left(
\alpha_1 \beta- \alpha \beta_1 -\alpha \beta - \alpha_1 \beta_1 \right) \right]
\end{eqnarray}
where we have omitted the common color factor. The second contribution is
evidently ($-$) the first after the replacements $\alpha 
\longleftrightarrow \beta$ and $\alpha_1 \longleftrightarrow \beta_1$. 
For the soft contributions of the three loop diagrams $s_5$, $s_6$, $s_9$, 
$s_{10}$, $s_{11}$ and $s_{12}$ where the antisymmetry is removed by the
additional gluon, we find it technically more convenient to choose 
$\left\{ k_2, p_1 \right\}$ for the Sudakov decomposition for the genuine
DL-corrections listed in appendix \ref{sec:C3l}. 
A careful treatment should in any case give the same result, however.
\begin{center}
\begin{figure}[t]
\centering
\epsfig{file=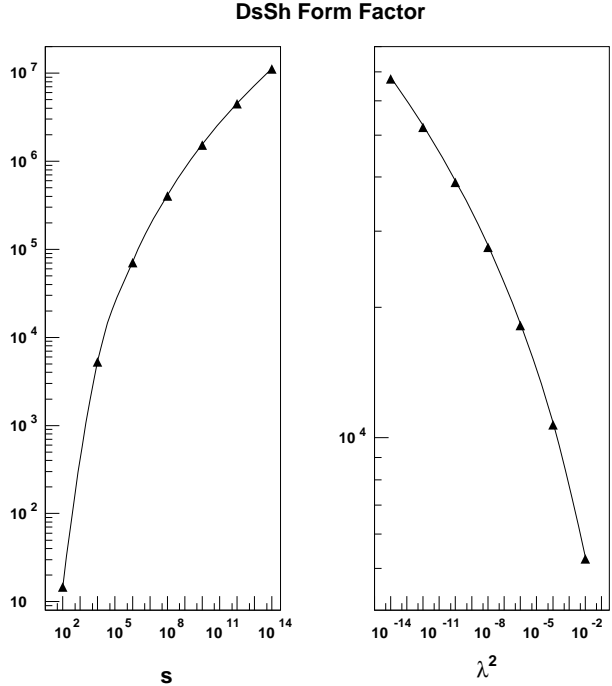,height=10cm}
\caption{A comparison of the third term of the series expansion in Eq.
\ref{eq:Chsff} (line) with a double soft and single hard form factor (DsSh)
and the explicit three loop results given in
appendix \ref{sec:hss} (triangles).
The plot on the left keeps $\lambda^2=10^{-2}$ fixed, the one on the right
has $s=(100$ GeV$)^2$ for each data point. For the three loop
results we used $10^6$ evaluations for each of the $100$ iterations and
$m_b = 4.5$ GeV was used in both plots. The absolute values are plotted
and the agreement is below the percent
level. Color factors and the coupling constant are omitted.}
\label{fig:C3lhss}
\end{figure}
\end{center}

Fig. \ref{fig:C3lhss} shows that the three loop contribution of the sum of
diagrams $s_1+...+s_{12}$ is indeed equal to the product of
the one-loop hard form factor in Eq. \ref{eq:1lhff} multiplied by the second term
in the Sudakov exponential of Eq. \ref{eq:softff}. The leading logarithmic
behavior is shown to agree for both $\lambda \to 0$ as well as for
$s \to \infty$, keeping $m=m_b$ fixed. The relative difference actually
decreases with increasing values of the large logarithms.

It is therefore more than suggestive that the structure which prevails for the
one-, two- and three-loop DL-contributions exponentiates to all orders 
analogously to that of topology ${\cal B}$ in Eq. \ref{eq:Bhsff}, namely that

\begin{equation}
{\cal F}^{hs}_{\cal C} =
{\cal F} \;\; 
_2F_2 (1,1;2,\frac{3}{2}; \frac{C_A}{4 C_F}  
{\cal F} ) \; \left[  \; \exp \left( {\cal F}_{\cal A} \right) -1
\right] \label{eq:Chsff}
\end{equation}

\section{Summary}

In this paper we have derived a novel series of hard double logarithms
in polarized $\gamma \gamma \left( J_z=0 \right)$ collisions to all orders. The
resulting series is given by a confluent hypergeometric function $_2F_2$ which
possesses a $\log \left( \frac{\alpha_s}{\pi} \log^2 \frac{m^2}{s} \right)$
high energy limit. Taking into account the $\frac{m}{\sqrt{s}}$ suppression
contained in the Born amplitude, the new corrections are well behaved as
$s \longrightarrow \infty$. The series expansion agrees with the known one and
two loop results as well as our explicit three loop calculation.

It was furthermore shown that at three loops all additional doubly logarithmic
corrections are described by the exponentiation of Sudakov logarithms
at each order in the expansion of the new hard form factor. This behavior
is already present at one and two loops and can thus safely be extrapolated
to all orders.

In summary we list the complete virtual DL-form factor contribution to the process 
$\gamma + \gamma \left( J_z=0 \right) \longrightarrow q + \overline{q}$:
\begin{eqnarray}
{\cal M}_{DL} &=& {\cal M}_{Born} \left\{ \exp \left( {\cal F}_{\cal A} \right) +
{\cal F} \;\; 
_2F_2 (1,1;2,\frac{3}{2}; \frac{1}{2} 
{\cal F} ) +
2 \; {\cal F} \;\; 
_2F_2 (1,1;2,\frac{3}{2}; \frac{C_A}{4 C_F} 
{\cal F} ) \right. \nonumber \\
&& \;\;\;\;\;\;\;\;\;\;\;\;\;\; + {\cal F} \;\; 
_2F_2 (1,1;2,\frac{3}{2}; \frac{1}{2} 
{\cal F} ) \; \left[  \; \exp \left( {\cal F}_{\cal A} \right) -1 \right] 
\nonumber \\
&& \left. \;\;\;\;\;\;\;\;\;\;\;\;\;\; + 2 \; {\cal F} \;\; 
_2F_2 (1,1;2,\frac{3}{2}; \frac{C_A}{4 C_F}  
{\cal F} ) \; \left[  \; 
\exp \left( {\cal F}_{\cal A} \right) -1
\right] \right\} \\
&=& {\cal M}_{Born} \left\{ 1 +
{\cal F} \;\; 
_2F_2 (1,1;2,\frac{3}{2}; \frac{1}{2} 
{\cal F} ) +
2 \; {\cal F} \;\; 
_2F_2 (1,1;2,\frac{3}{2}; \frac{C_A}{4 C_F} 
{\cal F} ) \right\} 
\exp \left( {\cal F}_{\cal A} \right) \nonumber 
\label{eq:DL}
\end{eqnarray}
where ${\cal F}_{\cal A}$ and ${\cal F}$ denote the soft and hard one-loop
form factors of Eqs. \ref{eq:sff} and \ref{eq:1lhff} respectively.
It should be pointed out that the contributions containing the soft
form factor ${\cal F}_{\cal A}$ cancel in a physical cross section with
the infrared divergent contributions from the emission of real soft
gluons. For realistic future collider applications Fig. \ref{fig:rel} 
demonstrated that the higher than second order contributions are important
if an accuracy in the percentile regime could be reached. 
A more thorough study of the phenomenological implications of this work 
including realistic
experimental cuts is in progress \cite{ms}.

\vspace{0.5cm}
\noindent{\bf Acknowledgements}\\ 
The Authors would like to thank M. W\"usthoff for valuable discussions
and M. Heyssler for helpful advice on using ``Paw''. 
This work was supported in part by the EU Fourth Framework Programme `Training and Mobility of 
Researchers', Network `Quantum Chromodynamics and the Deep Structure of Elementary Particles', 
contract FMRX-CT98-0194 (DG 12 - MIHT).  
 
\appendix

\section{Three Loop Results in the DL-Approximation} \label{sec:C3l}

\subsection{Hard and Soft Corrections} \label{sec:has}

\begin{center}
\begin{figure}
\centering
\epsfig{file=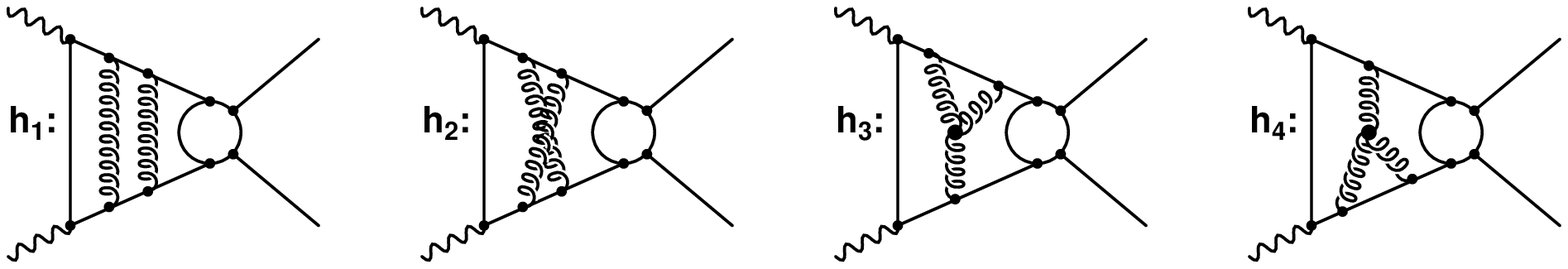,width=12cm}
\vspace{1.0cm} \\
\epsfig{file=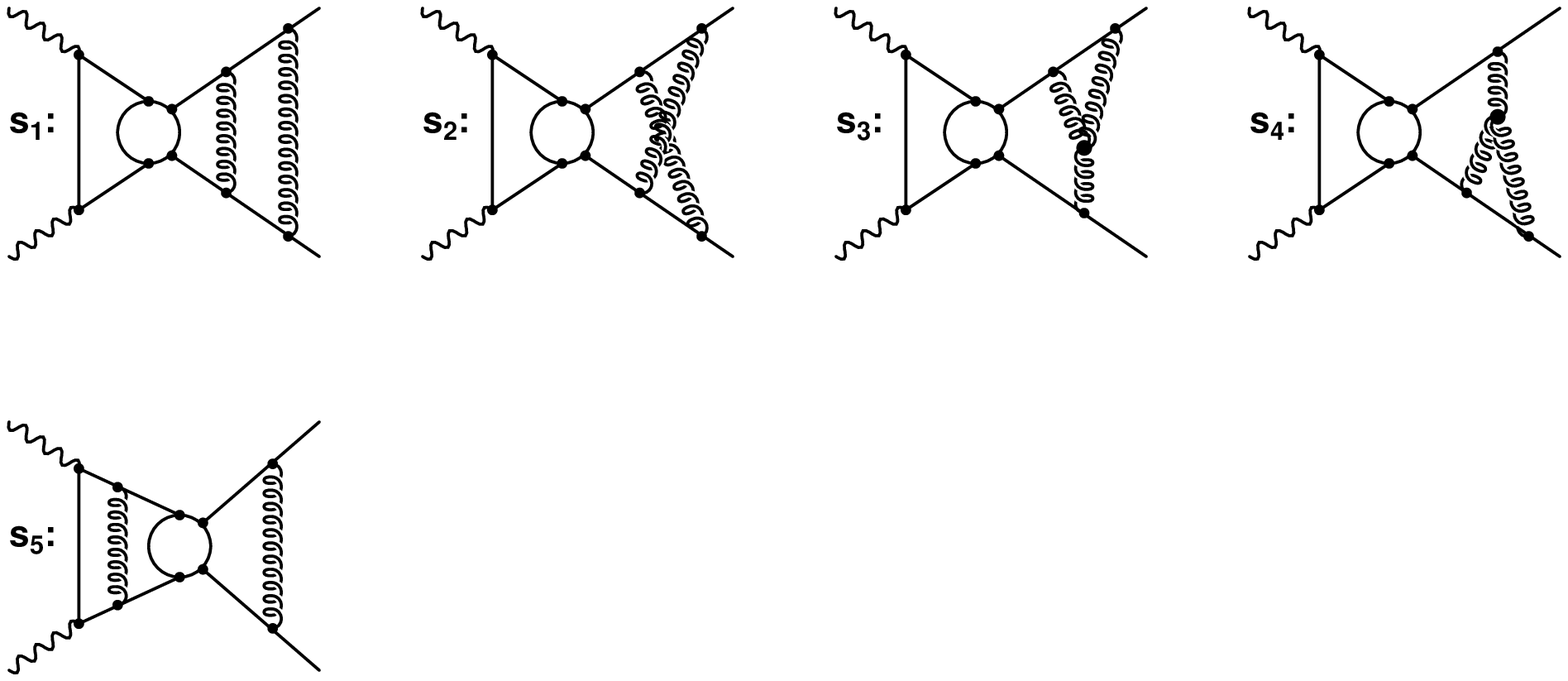,width=12cm}
\caption{The three loop Feynman diagrams contributing to the purely hard
 (h) and soft (s) form factors of the topology ${\cal B}$. 
 The sum of all soft terms exponentiates from each order of the hard process.
 The remaining diagrams cancel pairwise as described in section \ref{sec:res}.}
\label{fig:B3l}
\end{figure}
\end{center}

In this appendix we give explicit three loop results for the $\Theta_k$ terms
of Eq. \ref{eq:NSudBC} of topology ${\cal C}$. 
The particular contributions are listed in Fig.
\ref{fig:C3l} with the corresponding notation. For the three loop 
contribution of topology ${\cal B}$ in Fig. \ref{fig:B3l} the left and right
side of the hard blob factorize for the graphs $s_{1},...,s_{5}$. On a 
technical level they are therefore effectively two 
loop corrections. See Ref. \cite{fkm} for explicit results. 

In the following we distinguish between corrections with one and those with
two purely hard form factors. We also omit color factors at this point, as
the exponentiation of the soft gluons ensures that all ``higher'' order
$C_A$ contributions are canceled by purely non-Abelian corrections
\cite{ct,ds,Sen}.
The correct color factors of the sum of the graphs given in Figs. 
\ref{fig:C3l}, \ref{fig:C3lsna} and \ref{fig:C3lz} are contained in the
results of section \ref{sec:res}.

\subsubsection{Double Soft Single Hard Contributions} \label{sec:hss}

The following results for the $\Theta_k$ correspond to the respective
graphs of Fig. \ref{fig:C3l} with two potentially soft gluons.
The Sudakov parametrization for the three soft momenta (as indicated in
the figure for $s_1$) was chosen to be 
$\left\{ k_2, \left( p_2- \frac{m^2}{s} k_2 \right)
\right\}$ for the soft fermion line $l$ in all graphs,
$\left\{ \left( p_1 - \frac{m^2}{s} p_2 \right), \left( p_2- \frac{m^2}{s} 
p_1 \right) \right\}$ for $l_1$ in $s_1,s_2$ and for $l_2$ in 
$s_1,...,s_6$ and $\left\{ k_2, \left( p_1- \frac{m^2}{s} k_2 \right) \right\}$
for $l_1$ in $s_3,...s_{12}$ and for $l_2$ in $s_7,...s_{12}$.

\begin{eqnarray}
\Theta_{s_1} &\sim& -\Theta \left( \alpha \beta - \frac{m^2}{s} \right)
\Theta \left( \alpha_1 \beta_1 - \frac{\lambda^2}{s} \right)
\Theta \left( \alpha_1 - \frac{m^2}{s} \beta_1 \right)
\Theta \left( \beta_1 - \frac{m^2}{s} \alpha_1 \right) \nonumber \\
&& \Theta \left( \alpha_2 - \frac{m^2}{s} \beta_2 \right)
\Theta \left( \beta_2 - \frac{m^2}{s} \alpha_2 \right)
\Theta \left( \alpha_2 - \alpha_1 \right) \Theta \left( \beta_2 - \beta_1 \right)
\Theta \left( \beta - \beta_2 \right) \label{eq:s1} \\
\Theta_{s_2} &\sim& -\Theta \left( \alpha \beta - \frac{m^2}{s} \right)
\Theta \left( \alpha_1 \beta_1 - \frac{\lambda^2}{s} \right)
\Theta \left( \alpha_2 \beta_2 - \frac{\lambda^2}{s} \right)
\Theta \left( \alpha_1 - \frac{m^2}{s} \beta_1 \right) \nonumber \\
&& \Theta \left( \beta_1 - \frac{m^2}{s} \alpha_2 \right) 
\Theta \left( \alpha_2 - \frac{m^2}{s} \beta_1 \right) 
\Theta \left( \beta_2 - \frac{m^2}{s} \alpha_2 \right)
\Theta \left( \alpha_2 - \alpha_1 \right) \Theta \left( \beta_1 - \beta_2 
\right) \nonumber \\
&& \Theta \left( \beta - \beta_1 \right) \label{eq:s2} \\
\Theta_{s_3} &\sim& -\Theta \left( \alpha \beta - \frac{m^2}{s} \right)
\Theta \left( \alpha_1 \beta_1 - \frac{\lambda^2}{s} \right)
\Theta \left( \alpha_2 \beta_2 - \frac{\lambda^2}{s} \right)
\Theta \left( \alpha_1 - \frac{m^2}{s} \beta_1 \right) \nonumber \\
&& \!\!\!\! \Theta \left( \beta_1 - \frac{m^2}{s} \alpha_1 \right) 
\Theta \left( \beta_2 - \frac{m^2}{s} \left( \alpha_2 + \beta_1 \right) \right)
\Theta \left( \alpha_2 - \alpha \right) \Theta \left( \beta_2 - \alpha_1 \right)
\Theta \left( \beta - \beta_1 \right) \label{eq:s3} \\
\Theta_{s_4} &\sim& -\Theta \left( \alpha \beta - \frac{m^2}{s} \right)
\Theta \left( \alpha_1 \beta_1 - \frac{\lambda^2}{s} \right)
\Theta \left( \alpha_2 \beta_2 - \frac{\lambda^2}{s} \right)
\Theta \left( \beta_2 - \frac{m^2}{s} \alpha_2 \right) \nonumber \\
&& \!\!\!\!  \Theta \left( \beta_1 - \frac{m^2}{s} \alpha_1 \right) 
\Theta \left( \alpha_1 - \frac{m^2}{s} \left( \alpha_2 + \beta_1 \right) \right)
\Theta \left( \alpha_2 - \alpha \right) \Theta \left( \alpha_1 -\beta_2 \right)
\Theta \left( \beta - \beta_1 \right) \label{eq:s4} \\
\Theta_{s_5} &\sim& -\Theta \left( \alpha \beta - \frac{m^2}{s} \right)
\Theta \left( \alpha_1 \beta_1 - \frac{\lambda^2}{s} \right)
\Theta \left( \alpha_2 \beta_2 - \frac{\lambda^2}{s} \right)
\Theta \left( \beta_1 - \frac{m^2}{s} \alpha_1 \right) \nonumber \\
&& \Theta \left( \alpha_1 - \frac{m^2}{s} \beta_1 \right) 
\Theta \left( \alpha_2 - \alpha_1 \right) 
\Theta \left( \left( \alpha+\beta \right) \beta_2-\alpha \alpha_2-\alpha
\beta-\alpha_2 \beta_2 \right) \nonumber \\ && 
\Theta \left( \alpha_2 - \frac{m^2}{s} \left(\beta_1+\beta_2 \right) \right)
\left\{
\Theta \left( \alpha - \beta_2 \right)  
\Theta \left( \beta - \beta_1 \right) - 
\Theta \left( \beta -\beta_1 -\beta_2-\alpha_2 \right)
\right\} \label{eq:s5} \\
\Theta_{s_6} &\sim& -\Theta \left( \alpha \beta - \frac{m^2}{s} \right)
\Theta \left( \alpha_1 \beta_1 - \frac{\lambda^2}{s} \right)
\Theta \left( \alpha_2 \beta_2 - \frac{\lambda^2}{s} \right)
\Theta \left( \beta_1 - \frac{m^2}{s} \alpha_1 \right) \nonumber \\
&& \Theta \left( \beta_2 - \frac{m^2}{s} \alpha_2 \right)
\Theta \left( \alpha_1 - \beta_2 \right) 
\Theta \left( \left( \alpha+\beta \right) \alpha_2-\alpha \beta_2-\alpha
\beta-\alpha_2 \beta_2 \right) \nonumber \\ && 
\Theta \left( \alpha_1 - \frac{m^2}{s} \left(\beta_1+\alpha_2 \right) \right)
\left\{
\Theta \left( \alpha - \alpha_2 \right)  
\Theta \left( \beta - \beta_1 \right) - 
\Theta \left( \beta -\beta_1 -\beta_2-\alpha_2 \right)
\right\} \label{eq:s6} \\
\Theta_{s_7} &\sim& -\Theta \left( \alpha \beta - \frac{m^2}{s} \right)
\Theta \left( \alpha_1 \beta_1 - \frac{\lambda^2}{s} \right)
\Theta \left( \beta_1 - \frac{m^2}{s} \alpha_1 \right) \nonumber \\
&& \Theta \left( \beta_2 - \frac{m^2}{s} \alpha_2 \right)
\Theta \left( \alpha_2 - \alpha_1 \right) \Theta \left( \beta_2 - \beta_1 \right)
\Theta \left( \alpha_1 - \alpha \right) \label{eq:s7} \\
\Theta_{s_8} &\sim& -\Theta \left( \alpha \beta - \frac{m^2}{s} \right)
\Theta \left( \alpha_1 \beta_1 - \frac{\lambda^2}{s} \right)
\Theta \left( \alpha_2 \beta_2 - \frac{\lambda^2}{s} \right)
\Theta \left( \beta_1 - \frac{m^2}{s} \alpha_1 \right) \nonumber \\
&& \Theta \left( \beta_2 - \frac{m^2}{s} \alpha_1 \right)
\Theta \left( \alpha_2 - \alpha \right) \Theta \left( \beta_2 - \beta_1 \right)
\Theta \left( \alpha_1 - \alpha_2 \right) \label{eq:s8} \\
\Theta_{s_9} &\sim& -\Theta \left( \alpha \beta - \frac{m^2}{s} \right)
\Theta \left( \alpha_1 \beta_1 - \frac{\lambda^2}{s} \right)
\Theta \left( \alpha_2 \beta_2 - \frac{\lambda^2}{s} \right)
\Theta \left( \beta_2 - \frac{m^2}{s} \alpha_2 \right) \nonumber \\
&& \Theta \left( \beta_1 - \beta_2 \right) 
\Theta \left( \beta_1 - \frac{m^2}{s} \left(\alpha_1+\alpha_2 \right) \right)
\Theta \left( \left( \alpha+\beta \right) \alpha_2-\alpha \beta_2-\alpha
\beta-\alpha_2 \beta_2 \right) 
\nonumber \\ &&
\Theta \left( \alpha_1 - \alpha \right)
\left\{ \Theta \left( \alpha - \alpha_2 \right) 
- 
\Theta \left( \beta -\beta_1 -\alpha_1 \right)
\right\} \label{eq:s9} \\
\Theta_{s_{10}} &\sim& -\Theta \left( \alpha \beta - \frac{m^2}{s} \right)
\Theta \left( \alpha_1 \beta_1 - \frac{\lambda^2}{s} \right)
\Theta \left( \alpha_2 \beta_2 - \frac{\lambda^2}{s} \right)
\Theta \left( \beta_1 - \frac{m^2}{s} \alpha_1 \right)
\Theta \left( \beta_2 - \beta_1 \right) \nonumber \\ 
&& \Theta \left( \beta_2 - \frac{m^2}{s} \left(\alpha_1+\alpha_2 \right) \right)
\Theta \left( \left( \alpha+\beta \right) \alpha_2-\alpha \beta_2-\alpha
\beta-\alpha_2 \beta_2 \right)  
\nonumber \\ && \Theta \left( \alpha_1 - \alpha \right) 
\left\{ \Theta \left( \alpha - \alpha_2 \right) 
- \Theta \left( \beta -\beta_2 -\alpha_2 \right)
\right\} \label{eq:s10} \\
\Theta_{s_{11}} &\sim& -\Theta \left( \alpha \beta - \frac{m^2}{s} \right)
\Theta \left( \alpha_1 \beta_1 - \frac{\lambda^2}{s} \right)
\Theta \left( \alpha_2 \beta_2 - \frac{\lambda^2}{s} \right)
\Theta \left( \beta_1 - \frac{m^2}{s} \alpha_1 \right)
\Theta \left( \beta_2 - \beta_1 \right) \nonumber \\ 
&& \Theta \left( \beta_2 - \frac{m^2}{s} \left(\alpha_1+\alpha_2 \right) 
\right) \nonumber \\ &&
\left\{ \Theta \left( \beta - \beta_1 -\alpha_1 -\beta_2 -\alpha_2 \right)
\Theta \left( \left( \alpha+\beta \right) \alpha_2-\alpha \beta_2
-\alpha \beta-\alpha_2 \beta_2 \right) \right. \nonumber \\
&& \Theta \left( \left( \alpha+\beta \right) \alpha_1- \left( \alpha+
\beta \right) \alpha_2 - \alpha \beta_1
-\alpha_1 \beta_2-\alpha_2 \beta_1 -\alpha_1 \beta_1 \right) 
\;\; - \nonumber \\ && 
\Theta \left( \alpha - \alpha_2 \right) 
\Theta \left( \beta -\beta_1 -
\alpha_1 \right) 
\Theta \left( \left( \alpha+\beta \right) \alpha_1-\alpha \beta_1
-\alpha \beta-\alpha_1 \beta_1 \right) \nonumber \\ &&
\Theta \left( \left( \alpha+\beta \right) \alpha_2-\alpha \beta_2
-\alpha \beta-\alpha_2 \beta_2 \right) \;\; + \nonumber \\ &&  
\Theta \left( \alpha -
\alpha_1 -\alpha_2 \right) 
\Theta \left( \left( \alpha+
\beta \right) \alpha_2- \left( \alpha+ \beta \right) \alpha_1 - 
\alpha \beta_2 -\alpha_1 \beta_2-\alpha_2 \beta_1 -\alpha_2 \beta_2 \right)
\nonumber \\
&& \left. \Theta \left( \left( \alpha+\beta \right) \alpha_1-\alpha \beta_1
-\alpha \beta-\alpha_1 \beta_1 \right) \right\} \label{eq:s11} \\
\Theta_{s_{12}} &\sim& -\Theta \left( \alpha \beta - \frac{m^2}{s} \right)
\Theta \left( \alpha_1 \beta_1 - \frac{\lambda^2}{s} \right)
\Theta \left( \alpha_2 \beta_2 - \frac{\lambda^2}{s} \right)
\Theta \left( \beta_2 - \frac{m^2}{s} \alpha_2 \right)
\Theta \left( \beta_1 - \beta_2 \right) \nonumber \\ 
&& \Theta \left( \beta_1 - \frac{m^2}{s} \left(\alpha_1+\alpha_2 \right) 
\right) \nonumber \\ &&
\left\{ \Theta \left( \beta - \beta_1 -\alpha_1 -\beta_2 -\alpha_2 \right)
\Theta \left( \left( \alpha+\beta \right) \alpha_1-\alpha \beta_1
-\alpha \beta-\alpha_1 \beta_1 \right) \right. \nonumber \\
&& \Theta \left( \left( \alpha+\beta \right) \alpha_2- \left( \alpha+
\beta \right) \alpha_1 - \alpha \beta_2
-\alpha_1 \beta_2-\alpha_2 \beta_1 -\alpha_2 \beta_2 \right) 
\;\; - \nonumber \\ && 
\Theta \left( \alpha - \alpha_1 \right)
\Theta \left( \beta -\beta_2 -
\alpha_2 \right) 
\Theta \left( \left( \alpha+\beta \right) \alpha_1-\alpha \beta_1
-\alpha \beta-\alpha_1 \beta_1 \right) \nonumber \\ &&
\Theta \left( \left( \alpha+\beta \right) \alpha_2-\alpha \beta_2
-\alpha \beta-\alpha_2 \beta_2 \right) \;\; + \nonumber \\ &&  
\Theta \left( \alpha -
\alpha_1 -\alpha_2 \right) 
\Theta \left( \left( \alpha+
\beta \right) \alpha_1- \left( \alpha+ \beta \right) \alpha_2 - 
\alpha \beta_1 -\alpha_1 \beta_2-\alpha_2 \beta_1 -\alpha_1 \beta_1 \right)
\nonumber \\
&& \left. \Theta \left( \left( \alpha+\beta \right) \alpha_2-\alpha \beta_2
-\alpha \beta-\alpha_2 \beta_2 \right) \right\} \label{eq:s12}
\end{eqnarray}

\subsubsection{Double Hard Single Soft Contributions} \label{sec:hhs}

\begin{center}
\begin{figure}
\centering
\epsfig{file=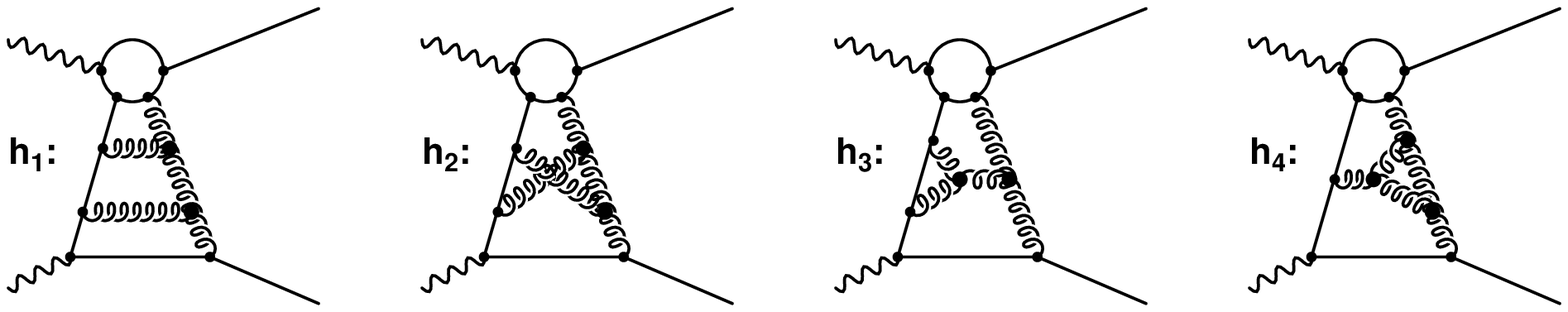,width=12cm}
\vspace{0.7cm} \\
\epsfig{file=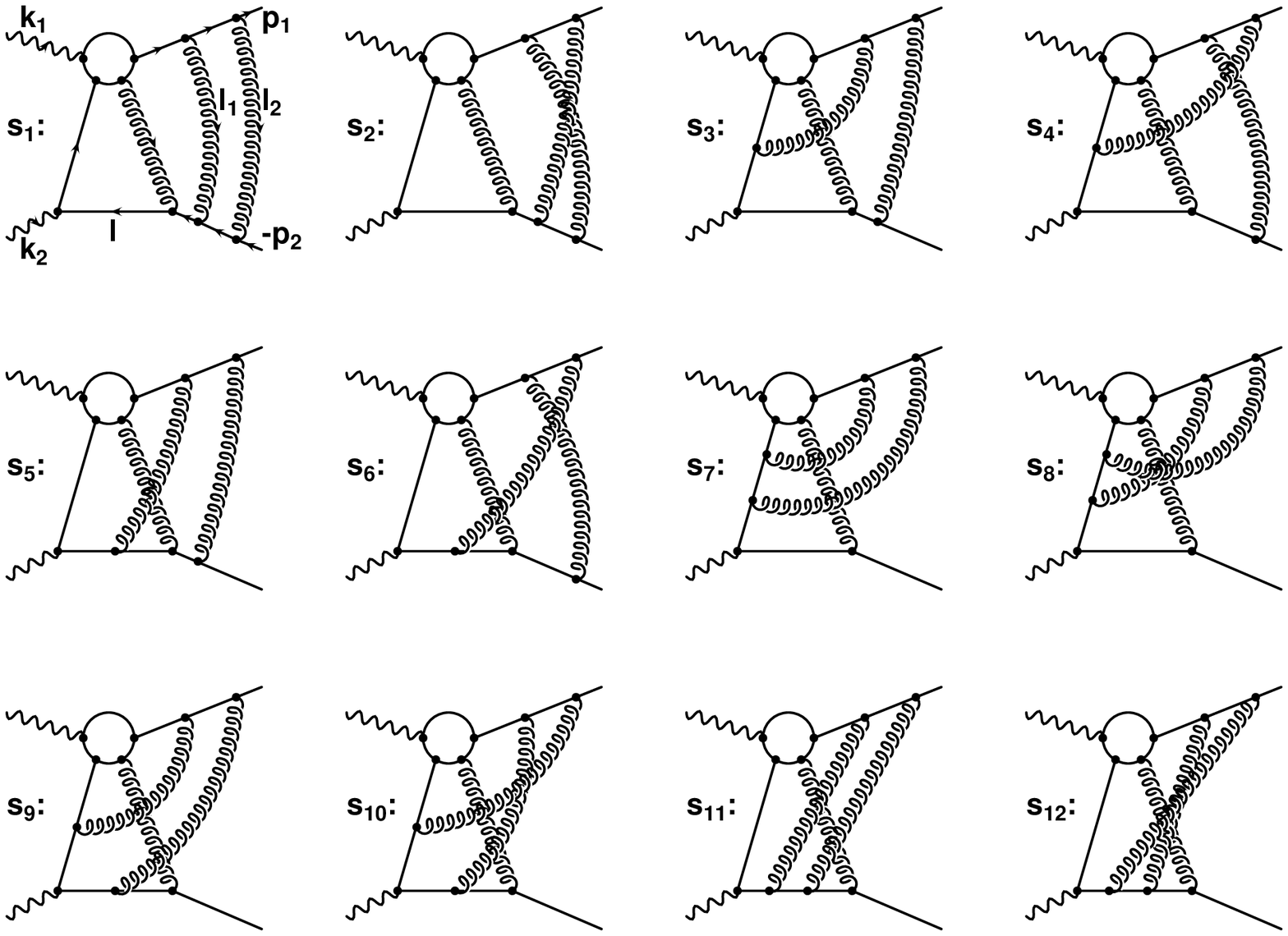,width=12cm}
\vspace{0.7cm} \\
\epsfig{file=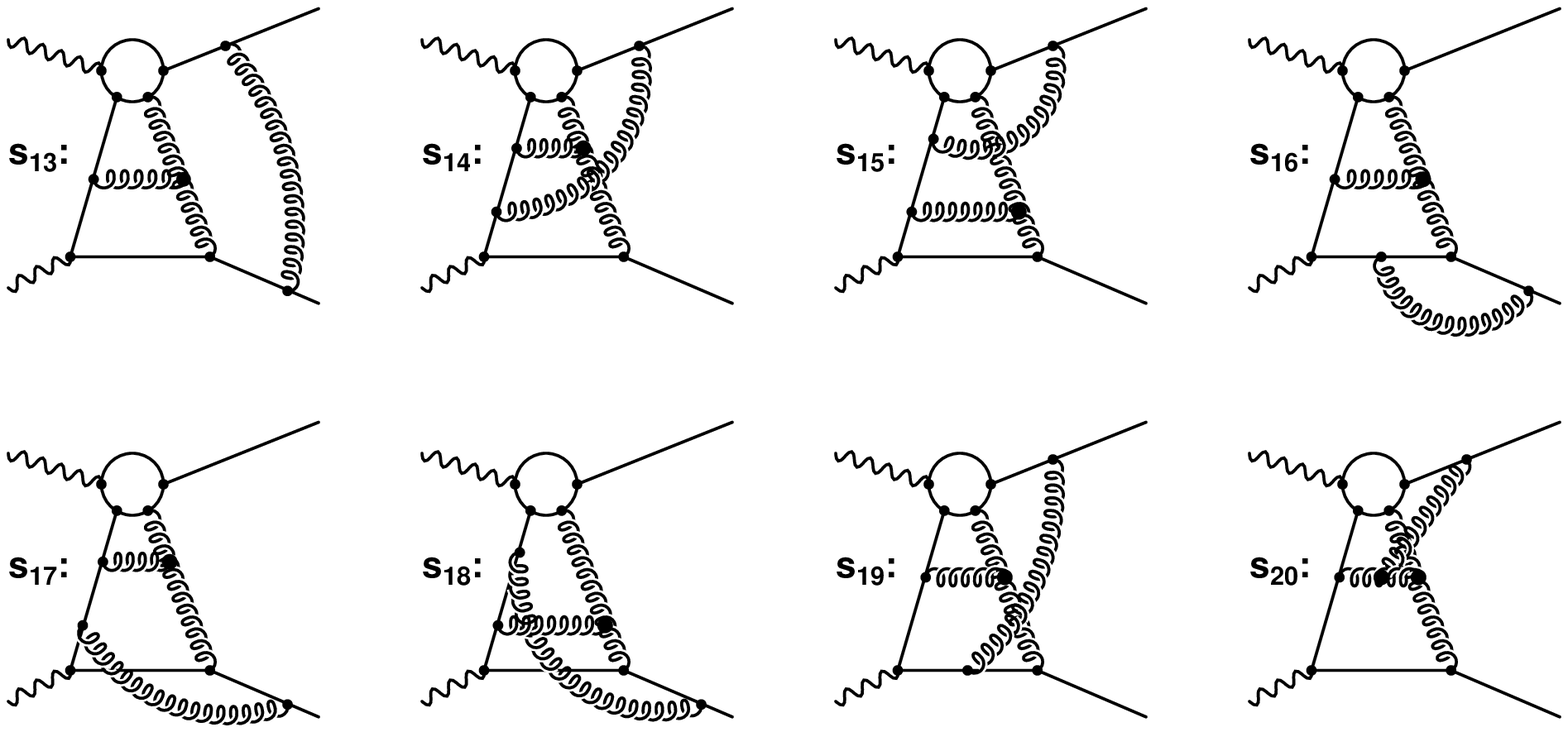,width=12cm}
\caption{The three loop Feynman diagrams contributing to the hard
 (h) and soft (s) form factors of the topology ${\cal C}$. 
 The sum of all soft terms exponentiates from each order of the hard process.}
\label{fig:C3l}
\end{figure}
\end{center}

Below we give the results for the $\Theta_k$ corresponding to the respective
graphs of Fig. \ref{fig:C3l} with one potentially soft gluon.
The Sudakov parametrization for the three soft momenta (as indicated in
the figure for $s_1$) was $\left\{ k_2, \left( p_2- \frac{m^2}{s} k_2 \right)
\right\}$ for the soft fermion line $l$ and for $l_1$ in all graphs and for
$l_2$ in $s_{16},...,s_{20}$, 
$\left\{ \left( p_1 - \frac{m^2}{s} p_2 \right), \left( p_2- \frac{m^2}{s} 
p_1 \right) \right\}$ for $l_2$ in $s_{13}$  
and $\left\{ k_2, \left( p_1- \frac{m^2}{s} k_2 \right) \right\}$
for $l_2$ in $s_{14},s_{15}$.

\begin{eqnarray}
\Theta_{s_{13}} &\sim& - \Theta \left( \alpha \beta - \frac{m^2}{s} \right)
\Theta \left( \alpha_2 \beta_2 - \frac{\lambda^2}{s} \right)
\Theta \left( \alpha_2 - \frac{m^2}{s} \beta_2 \right)
\Theta \left( \beta_2 - \frac{m^2}{s} \alpha_2 \right) \nonumber \\
&& \Theta \left( \beta - \beta_2 \right)
\Theta \left( \alpha_1 - \alpha \right) \Theta \left( \beta_1 - \beta \right)
\label{eq:s13} \\
\Theta_{s_{14}} &\sim& - \Theta \left( \alpha \beta - \frac{m^2}{s} \right)
\Theta \left( \alpha_2 \beta_2 - \frac{\lambda^2}{s} \right)
\Theta \left( \beta_2 - \frac{m^2}{s} \alpha_2 \right) \nonumber \\
&& \Theta \left( \alpha_1 - \alpha_2 \right)
\Theta \left( \alpha_2 - \alpha \right) \Theta \left( \beta_1 - \beta \right)
\label{eq:s14} \\
\Theta_{s_{15}} &\sim& -\Theta \left( \alpha \beta - \frac{m^2}{s} \right)
\Theta \left( \alpha_2 \beta_2 - \frac{\lambda^2}{s} \right)
\Theta \left( \beta_2 - \frac{m^2}{s} \alpha_2 \right) \nonumber \\
&& \Theta \left( \alpha_2 - \alpha_1 \right)
\Theta \left( \alpha_1 - \alpha \right) \Theta \left( \beta_1 - \beta \right)
\label{eq:s15} \\
\Theta_{s_{16}} &\sim& -\Theta \left( \alpha \beta - \frac{m^2}{s} \right)
\Theta \left( \alpha_2 \beta_2 - \frac{\lambda^2}{s} \right)
\Theta \left( \beta_2 - \frac{m^2}{s} \alpha_2 \right) \nonumber \\
&& \Theta \left( \alpha_1 - \alpha \right)
\Theta \left( \alpha_2 - \alpha \right) \Theta \left( \beta_1 - \beta \right)
\Theta \left( \beta - \beta_2 \right)
\label{eq:s16} \\
\Theta_{s_{17}} &\sim& \Theta \left( \alpha \beta - \frac{m^2}{s} \right)
\Theta \left( \alpha_2 \beta_2 - \frac{\lambda^2}{s} \right)
\Theta \left( \beta_2 - \frac{m^2}{s} \alpha_2 \right) \nonumber \\
&& \Theta \left( \alpha_1 - \alpha_2 \right)
\Theta \left( \alpha_2 - \alpha \right) \Theta \left( \beta_1 - \beta \right)
\Theta \left( \beta - \beta_2 \right)
\label{eq:s17} \\
\Theta_{s_{18}} &\sim& \Theta \left( \alpha \beta - \frac{m^2}{s} \right)
\Theta \left( \alpha_2 \beta_2 - \frac{\lambda^2}{s} \right)
\Theta \left( \beta_2 - \frac{m^2}{s} \alpha_2 \right) \nonumber \\
&& \Theta \left( \alpha_2 - \alpha_1 \right)
\Theta \left( \alpha_1 - \alpha \right) \Theta \left( \beta_1 - \beta \right)
\Theta \left( \beta - \beta_2 \right)
\label{eq:s18} \\
\Theta_{s_{19}} &\sim& 0 \label{eq:s19} \\
\Theta_{s_{20}} &\sim& 0 \label{eq:s20}
\end{eqnarray}

The last two diagrams vanish in the DL-approximation for the 
same reason as the two loop contribution
given in Fig. \ref{fig:2l}. The diagrams $s_{16},...,s_{18}$ cancel each other
after writing

\begin{equation}
\Theta \left( \alpha_1-\alpha \right) \Theta \left( \alpha_2-\alpha \right) 
= \Theta \left( \alpha_1-\alpha_2 \right) \Theta \left( \alpha_2-\alpha \right) 
+ \Theta \left( \alpha_2-\alpha_1 \right) \Theta \left( \alpha_1-\alpha \right) 
\end{equation}

\begin{center}
\begin{figure}
\centering
\epsfig{file=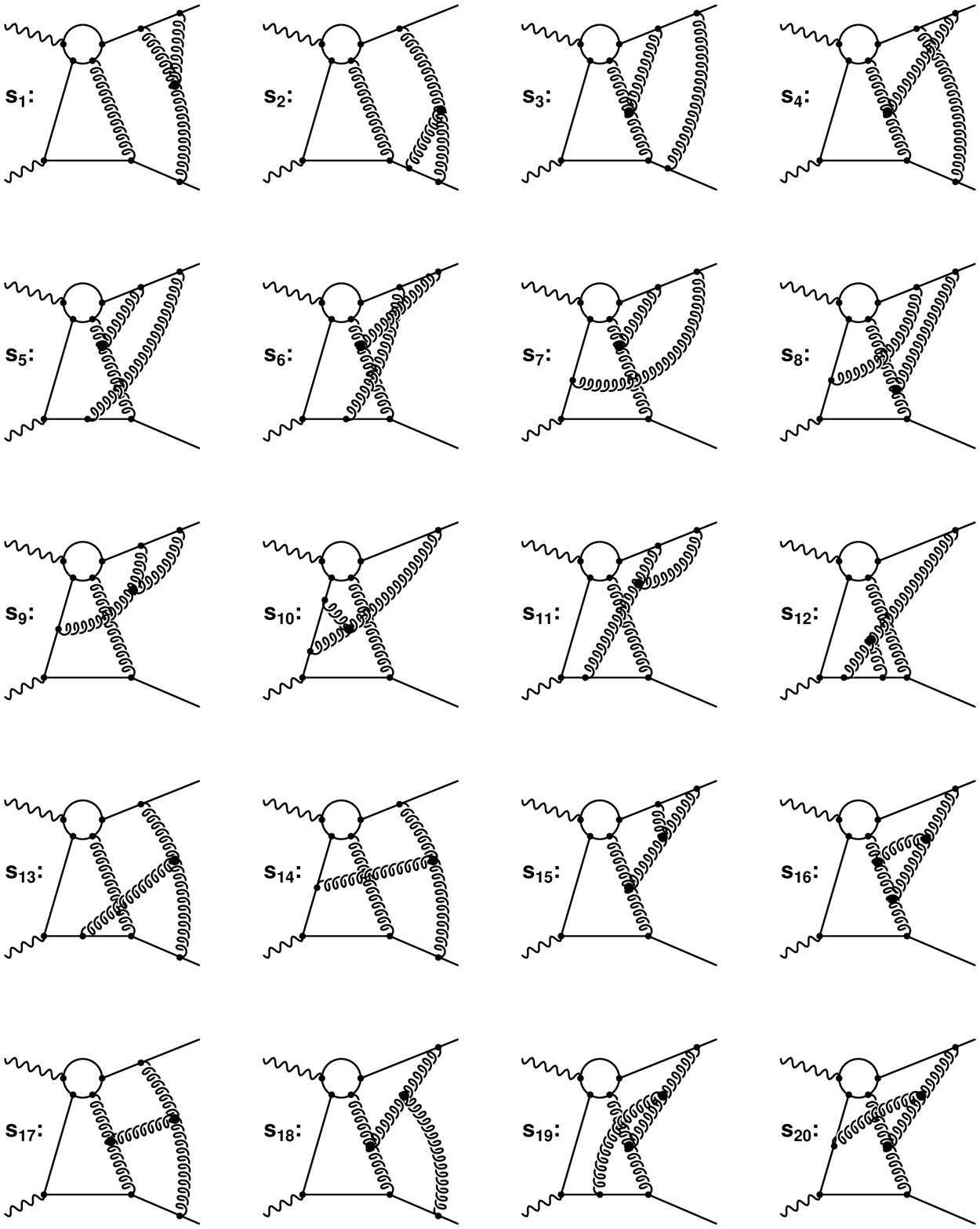,width=12cm}
\vspace{0.7cm} \\
\epsfig{file=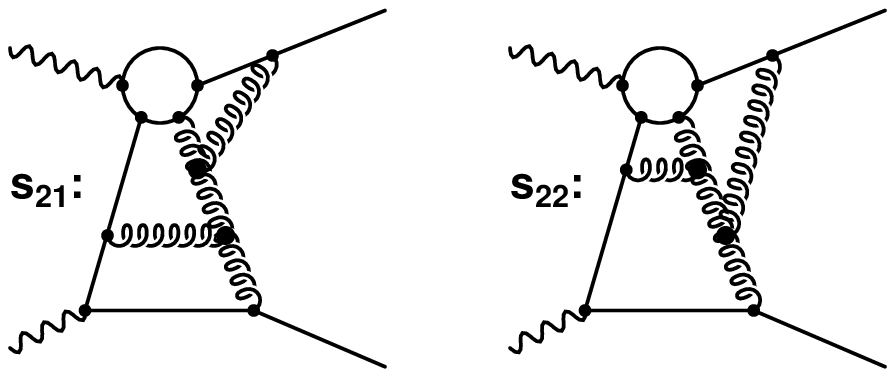,width=12cm}
\caption{The non-Abelian three loop Feynman diagrams 
 of the topology ${\cal C}$ which cancel the $C_A$ parts of color factors
 occurring in the soft contributions of Fig. \ref{fig:C3l}.}
\label{fig:C3lsna}
\end{figure}
\end{center}

Strictly speaking, the above cancellation takes place only for the $C_F^2C_A$
terms of amplitude $s_{18}$. The $C_FC_A^2$ term, absent in graphs $s_{16}$
and $s_{17}$ of Fig. \ref{fig:C3l}, is canceled by the analogous contribution
of graphs $s_{21}$ and $s_{22}$ of Fig. \ref{fig:C3lsna}. 

\begin{center}
\begin{figure}
\centering
\epsfig{file=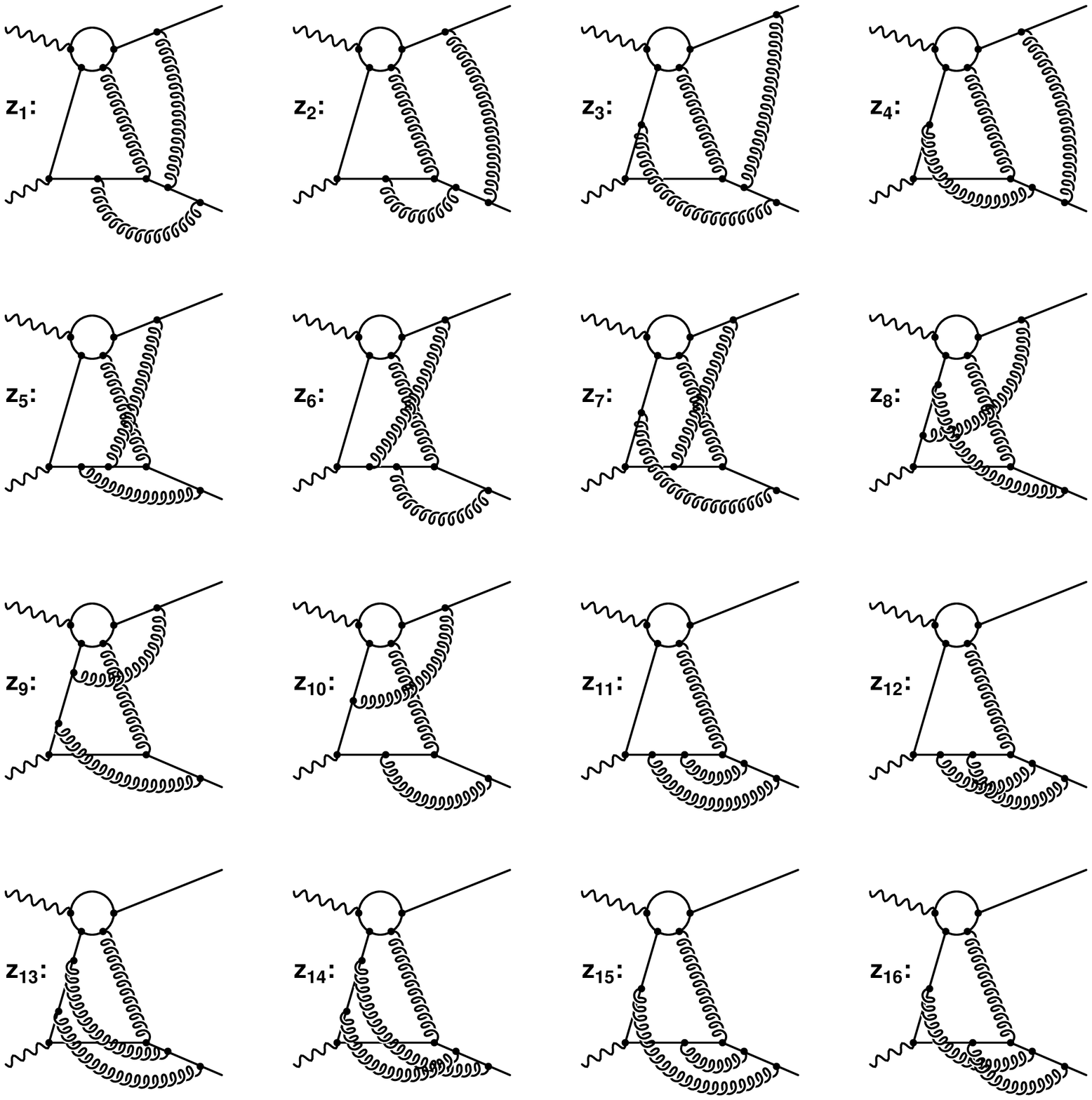,width=11cm}
\vspace{0.6cm} \\
\epsfig{file=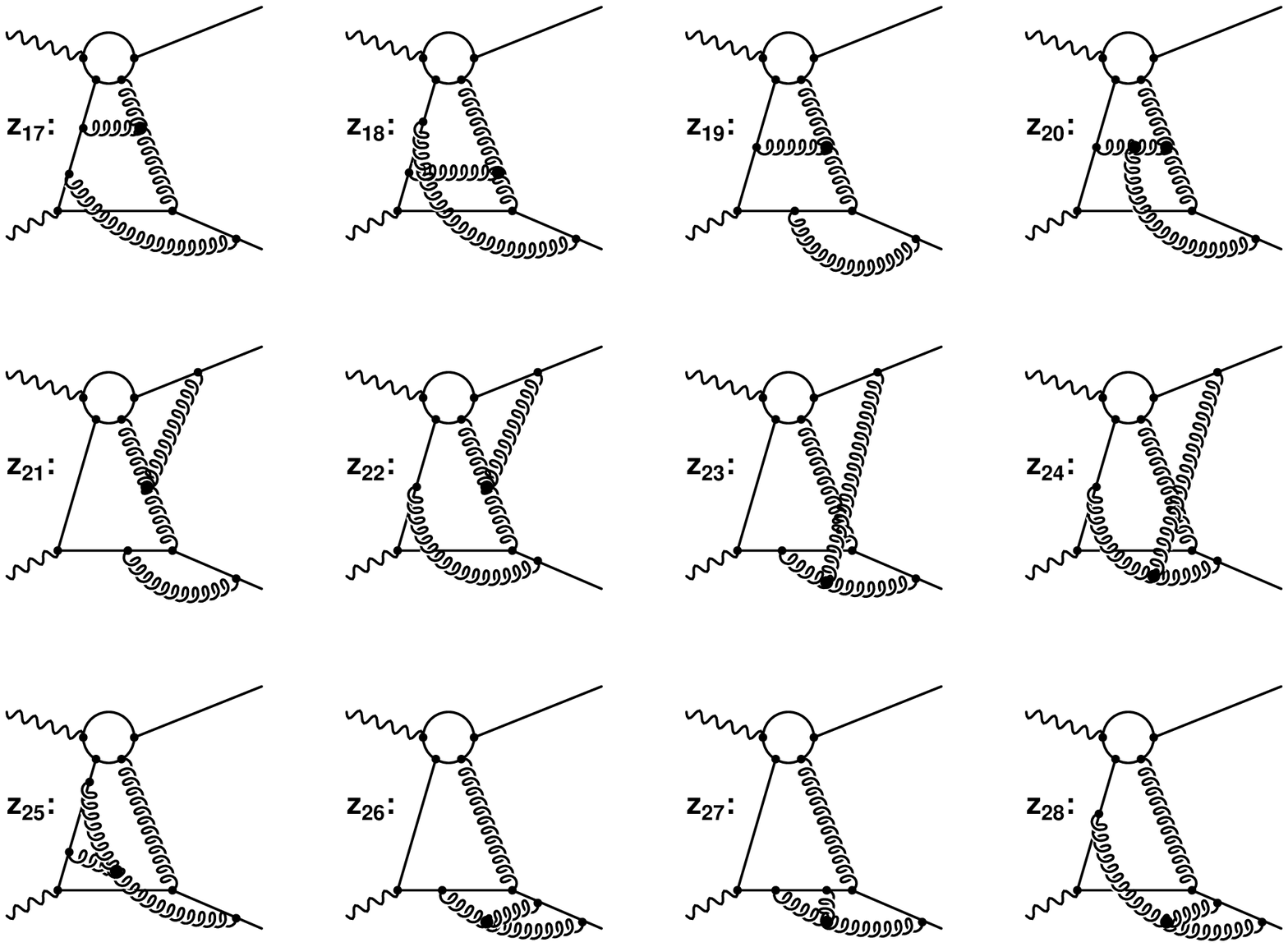,width=11cm}
\caption{The three loop Feynman diagrams 
 of the topology ${\cal C}$ which cancel either pairwise or in groups of three
 in the leading logarithmic approximation.} 
\label{fig:C3lz}
\end{figure}
\end{center}

In the same way the above mentioned cancellation of ``higher order'' $C_A$ terms
takes place with all non-Abelian graphs of Fig. \ref{fig:C3lsna}.
The remaining diagrams of topology ${\cal C}$ in Fig. \ref{fig:C3lz} all cancel
among themselves in groups of two or three.

\begin{center}
\begin{figure}
\centering
\epsfig{file=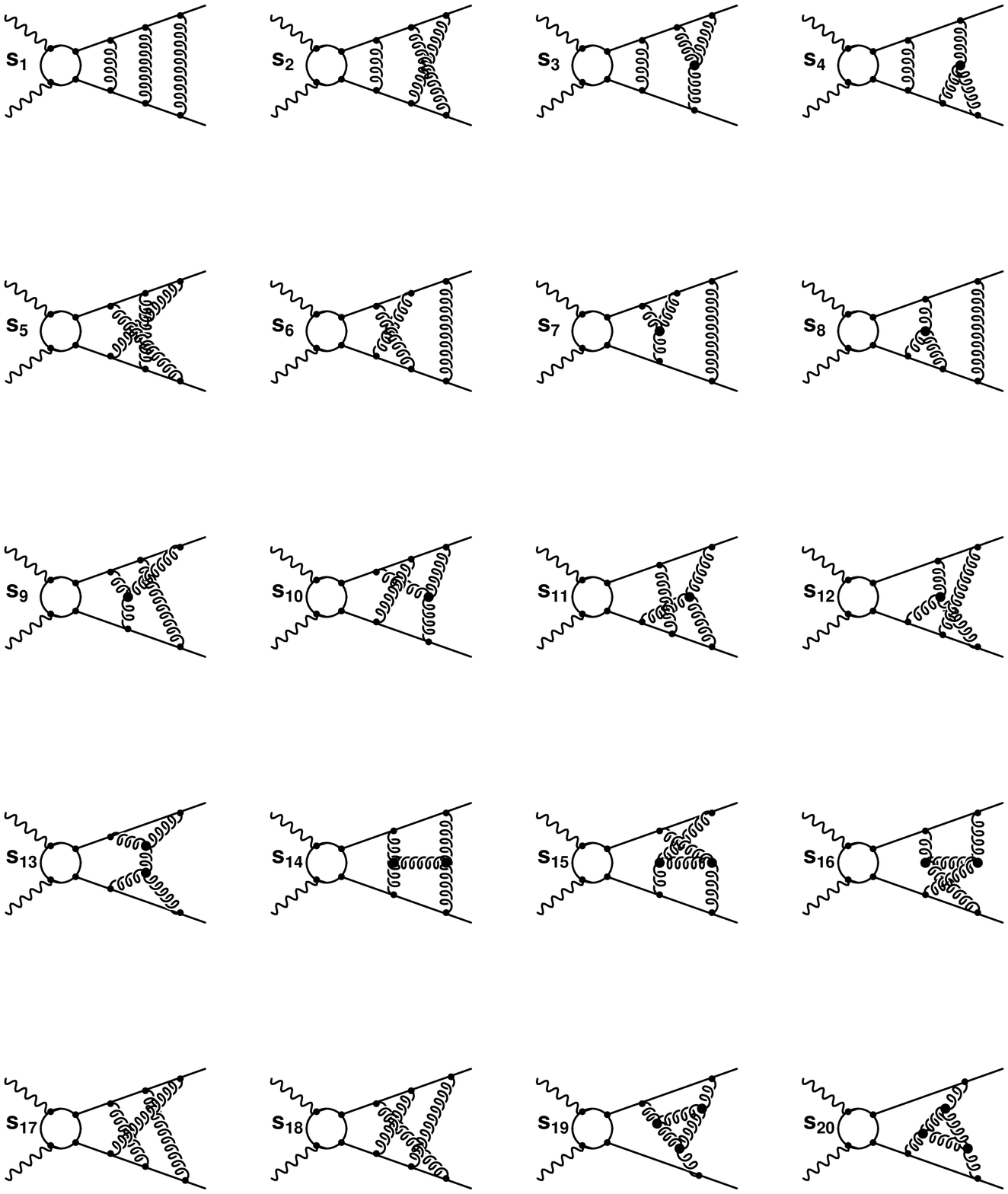,width=11cm}
\vspace{1.4cm} \\
\epsfig{file=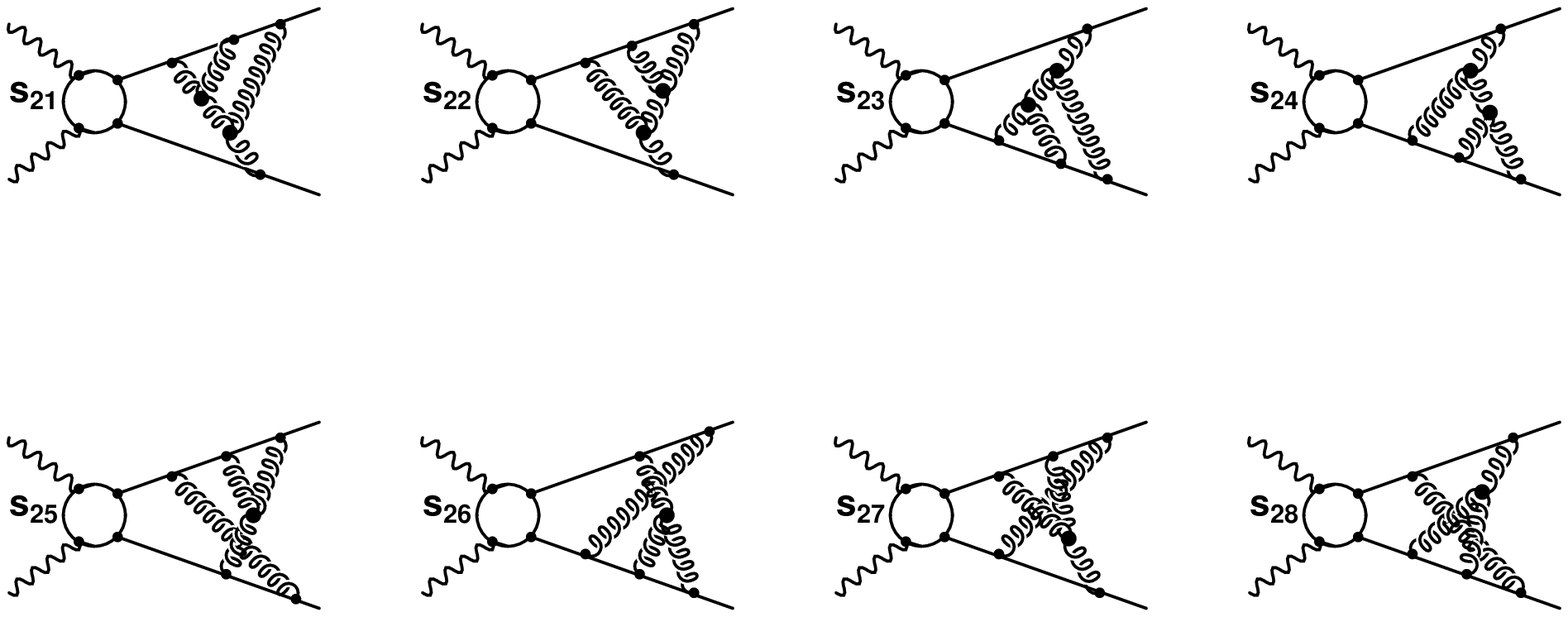,width=11cm}
\caption{The three loop Feynman diagrams contributing to the purely soft
 form factor of the topology ${\cal A}$. The sum of all terms exponentiates
 as usual with the non-Abelian diagrams canceling the $C_A$ parts of the
 crossed ``Abelian'' graphs.}
\label{fig:A3ls}
\end{figure}
\end{center}

\clearpage  
\begin{thebibliography}{999}  
 
\bibitem{gin1} I.F.~Ginzburg et al., 
Nucl. Inst. Meth. {\bf 205} (1983) 47. 

\bibitem{gin2} I.F.~Ginzburg et al., 
Nucl. Inst. Meth. {\bf 219} (1984) 5. 

\bibitem{tel} V.I.~Telnov, 
Nucl. Inst. Meth. {\bf A 355} (1995) 5. 

\bibitem{bkso} D.L.~Borden, V.A.~Khoze, J.~Ohnemus and W.J.~Stirling, 
Phys. Rev. {\bf D 50} (1994) 4499. 
  
\bibitem{jt1} G.~Jikia, A.~Takabladze, 
Phys. Rev. {\bf D 54} (1996) 2030. 

\bibitem{jt2} G.~Jikia, A.~Takabladze, 
Nucl. Inst. Meth. {\bf A 355} (1995) 81. 

\bibitem{fkm}  V.S.~Fadin, V.A.~Khoze and A.D.~Martin, 
 Phys. Rev. {\bf D 56} (1997) 484. 
 
\bibitem{ks} R.~Kleiss, W.J.~Stirling, 
Nucl. Phys. {\bf B 262} (1985) 235. 
  
\bibitem{lan} Landau, Lifshitz, 
Quantum Electrodynamics, Vol. 2, Pergamon Press, Oxford (1975). 

\bibitem{gglf} V.G.~Gorshkov, V.N.~Gribov, L.N.~Lipatov and G.V.~Frolov,
Sov. J. Nucl. Phys. {\bf 6} (1967) 95. 
  
\bibitem{kl} R.~Kirschner, L.N.~Lipatov,
Nucl. Phys. {\bf B 213} (1983) 122. 
  
\bibitem{ber} J.~Bartels, B.I.~Ermolaev and M.G.~Ryskin 
Z. Phys. {\bf C 70} (1996) 627. 

\bibitem{emr} B.I.~Ermolaev, S.I.~Manayenkov and M.G.~Ryskin 
Z. Phys. {\bf C 69} (1996) 259. 

\bibitem{Sud} V.V.~Sudakov, 
Sov. Phys. JETP {\bf 3} (1956) 65. 

\bibitem{ds} H.D.~Dahmen, F.~Steiner, 
Z. Phys. {\bf C 11} (1981) 247. 

\bibitem{ct} J.M.~Cornwall, G.~Tiktopoulos, 
Phys. Rev. {\bf D 13} (1976) 3370. 

\bibitem{Sen} A.~Sen, 
Phys. Rev. {\bf D 24} (1981) 3281. 

\bibitem{bu} V.V.~Belokurov, N.I.~Ussyukina, 
Phys. Lett. {\bf 94 B} (1980) 251. 

\bibitem{grad} Gradshteyn, Ryzhik, 
Table of Integrals, Series and Products, Academic Press (1994). 

\bibitem{veg} P.~Lepage, 
VEGAS, a Monte Carlo integrator, freely available, (c) (1975), Cornell 
University. 

\bibitem{ms} M.~Melles, W.J.~Stirling,
in preparation.

\end{thebibliography}
\end{document}